\documentclass[a4paper]{article}

\usepackage{mathtools}
\usepackage{amssymb}
\usepackage{bm}
\usepackage{chemformula}

\usepackage{siunitx}
\DeclareSIUnit\angstrom{\text {Å}}

\usepackage{caption}
\usepackage{subcaption}
\usepackage{stackengine}

\usepackage{booktabs}
\usepackage{ctable}

\usepackage[section]{placeins}

\usepackage[all]{nowidow}

\usepackage{hyperref}
\usepackage{xurl}

\usepackage[
  backend=biber,  
  style=numeric,       %
  sorting=none,        %
  giveninits=true,     %
  maxbibnames=10,      %
  url=false,
  eprint=false,
  date=year,
  uniquename=false,    %
  doi=true             %
]{biblatex}
\DeclareNameAlias{default}{family-given}
\DeclareDelimFormat{finalnamedelim}{\addspace\bibstring{and}\space} 

\DeclareFieldFormat[article]{journaltitle}{\mkbibemph{#1}}
\DeclareFieldFormat[article]{title}{#1}
\DeclareFieldFormat[article]{volume}{\mkbibbold{#1}}
\DeclareFieldFormat{pages}{#1}
\DeclareFieldFormat{doi}{#1}
\newbibmacro*{mydoi}{
  \iffieldundef{doi}
    {}
    {\printtext[parens]{doi\addcolon\space\href{https://doi.org/\thefield{doi}}{\nolinkurl{\thefield{doi}}}}}
}
\DeclareBibliographyDriver{article}{%
  \usebibmacro{bibindex}%
  \usebibmacro{begentry}%
  \usebibmacro{author}%
  \setunit{\addspace}%
  \printfield{year}%
  \setunit{\addspace}%
  \printfield{title}%
  \setunit{\addspace}%
  \printfield{journaltitle}%
  \setunit{\addspace}%
  \printfield{volume}%
  \setunit{\addspace}%
  \printfield{pages}%
  \setunit{\addspace}%
  \usebibmacro{mydoi}%
  \usebibmacro{finentry}%
}

\DeclareFieldFormat[book]{title}{\mkbibemph{#1}}
\DeclareBibliographyDriver{book}{%
  \usebibmacro{bibindex}%
  \usebibmacro{begentry}%
  \usebibmacro{author}%
  \setunit{\addspace}%
  \printfield{year}%
  \setunit{\addspace}%
  \printfield{title}%
  \setunit{\addspace}%
  \printtext[parens]{\printlist{publisher}}%
  \usebibmacro{finentry}%
}

\DeclareFieldFormat[inbook]{title}{#1}
\DeclareFieldFormat[inbook]{booktitle}{\mkbibemph{#1}}
\DeclareNameFormat{myeditor}{%
  \ifnum\value{listcount}=1
    \namepartfamily\space\namepartgiveni %
  \fi
  \ifnum\value{listcount}<\value{liststop}
    \addspace
  \fi
  \ifnum\value{listcount}=\value{liststop}
    \ifnum\value{liststop}>1
      \addspace et al.%
    \fi
  \fi
}
\DeclareFieldFormat[inbook]{pages}{p #1}
\DeclareBibliographyDriver{inbook}{%
  \usebibmacro{bibindex}%
  \usebibmacro{begentry}%
  \usebibmacro{author}%
  \setunit{\addspace}%
  \printfield{year}%
  \setunit{\addspace}%
  \printfield{title}%
  \setunit{\addspace}%
  \printfield{booktitle}%
  \setunit{\addspace}%
  \ifnameundef{editor}
  {}%
  {\printtext{ed\addspace}%
   \printnames[myeditor]{editor}}%
  \setunit{\addspace}%
  \printtext[parens]{\printlist{publisher}}%
  \setunit{\addspace}%
  \printfield{pages}%
  \usebibmacro{finentry}%
}

\DeclareFieldFormat[inproceedings]{title}{#1}
\DeclareFieldFormat[inproceedings]{booktitle}{\mkbibemph{#1}}
\DeclareBibliographyDriver{inproceedings}{%
  \usebibmacro{bibindex}%
  \usebibmacro{begentry}%
  \usebibmacro{author}%
  \setunit{\addspace}%
  \printfield{year}%
  \setunit{\addspace}%
  \printfield{title}%
  \setunit{\addspace}%
  \printfield{booktitle}%
  \setunit{\addspace}%
  \printfield{pages}%
  \setunit{\addspace}%
  \usebibmacro{mydoi}%
  \usebibmacro{finentry}%
}

\DeclareMathOperator{\Imag}{Im}
\DeclarePairedDelimiter\abs{\lvert}{\rvert} 
\DeclarePairedDelimiter\norm{\lVert}{\rVert} 
\DeclarePairedDelimiter{\bra}{\langle}{\rvert} 
\DeclarePairedDelimiter{\ket}{\lvert}{\rangle} 
\DeclarePairedDelimiterX\braket[2]{\langle}{\rangle}{#1\delimsize\vert\mathopen{}#2} 
\DeclarePairedDelimiterX\braketOP[3]{\langle}{\rangle}{#1\,\delimsize\vert\,\mathopen{}#2\,\delimsize\vert\,\mathopen{}#3} 

\newcommand{\orcid}[1]{\href{https://orcid.org/#1}{\includegraphics[width=8pt]{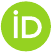}}}

\renewcommand{\title}[1]{{\exhyphenpenalty=10000\hyphenpenalty=10000 
 \fontsize{18}{21}\selectfont\noindent\raggedright
        \textsf{#1}\par}\suppressfloats[t]}

\renewcommand{\author}[1]{{\vspace{5mm}%
   \fontsize{10}{12}
      \raggedright \if@anonymous Author list removed for anonymity \else #1 \fi
	  \vspace{3mm}}}

\newcommand{\affil}[1]{{\fontsize{8}{10}\selectfont
       \raggedright \if@anonymous \phantom{#1} \else #1 \fi}
	   }

\newcommand{\email}[1]{\vspace*{12pt}{\fontsize{8}{10}\selectfont
       \raggedright {\bfseries E-mail:} \if@anonymous \phantom{#1} \else #1 \fi}
	  \vspace{3mm} }
	   
\newcommand{\keywords}[1]{{\fontsize{8}{10}\selectfont
       \raggedright {\bfseries Keywords:} #1}
	  \vspace{3mm} }

\newcommand{\note}[1]{{\fontsize{8}{10}\selectfont
       \raggedright #1}
	  }

\begin{document}

\title{First principles calculations of electric‑field‑driven topological phase transitions in silicene, germanene and stanene}

\author{Julián Antonio Villarreal Murúa$^{1*}$\orcid{0000-0002-9065-8943}, Pablo Roura-Bas$^{2,3}$\orcid{0000-0003-0758-4048}, Javier Daniel Fuhr$^{2,3}$\orcid{0000-0002-2640-4842} and Ricardo Faccio$^1$\orcid{0000-0003-1650-7677}}

\affil{$^1$Departamento de Experimentación y Teoría de la Estructura de la Materia y sus Aplicaciones, Facultad de Química, Universidad de la República, Montevideo, Uruguay}

\affil{$^2$División Física de Superficies, Centro Atómico Bariloche, Comisión Nacional de Energía Atómica, San Carlos de Bariloche, Argentina}

\affil{$^3$Instituto Balseiro, Universidad Nacional de Cuyo, San Carlos de Bariloche, Argentina}

\affil{$^*$Author to whom any correspondence should be addressed.}

\email{javillarreal@fq.edu.uy}


\note{This is the version of the article before editing, as submitted to \textit{Journal of Physics: Condensed Matter}. The Version of Record is available online at DOI: \href{https://doi.org/10.1088/1361-648X/ae9e75}{10.1088/1361-648X/ae9e75}.}

\begin{abstract}
The emergence of two-dimensional topological materials, particularly the group-14 monolayers known as silicene, germanene, and stanene has opened promising pathways for next-generation nanoelectronics and spintronics. Their buckled honeycomb structure and strong spin-orbit coupling allow for bandgap engineering via a perpendicular electric field, leading to topological phase transitions (TPTs) from non-trivial to trivial insulating states. However, precise determination of the critical electric field $E_z^{\text{cr}}$ at which these transitions occur remains challenging, with tight-binding models often underestimating these values. Here, we present a first-principles framework that combines density-functional theory (DFT), maximally localized Wannier functions, and evolution of the Wannier charge centers (WCC) to accurately characterize TPTs in silicene, germanene, and stanene through the $\mathbb{Z}_2$ topological invariant. In contrast to earlier work, at each electric-field strength we run fully self-consistent ab initio simulations to obtain the screened electronic structure, accounting for the material's dielectric response from both electrons and ions. From these converged results we construct a Wannier tight-binding Hamiltonian at each electric field strength, which then enables a gauge-invariant calculation of the $\mathbb{Z}_2$ topological invariant. This methodology yields significantly more accurate numerical predictions of $E_z^{\text{cr}}$---$\num{0.020}$ and $\qty{0.250}{V/\angstrom}$ for silicene and germanene, respectively---and provides deeper insight into the interplay between electronic structure and topological order. Compared to previous approaches, our framework delivers a marked quantitative improvement for predicting topological phase boundaries, essential for guiding the design of topological field-effect transistors and electrostatically controlled quantum devices based on two-dimensional materials.
\end{abstract}

\section{Introduction}
Since the beginning of the 21st century, the study of materials exhibiting topological properties has emerged as a technologically relevant field---especially for nanoelectronics \cite{Gilbert_2021, Culcer_2020}. In 2016, the Nobel Prize in Physics was awarded to Thouless, Haldane, and Kosterlitz for ``theoretical discoveries of topological phase transitions and topological phases of matter'' \cite{Haldane_2017, Kosterlitz_2017}. Shortly after the foundational theoretical predictions, the first experimental realization of a topological insulator—the quantum spin Hall state in HgTe quantum wells—was reported by König et al. \cite{Konig_2007}, providing a crucial experimental anchor for the field. This breakthrough demonstrated that topological electronic states could be engineered in solid-state systems and spurred intense research into other material platforms. Among the various kinds of topological materials are topological insulators---materials that, despite exhibiting a bandgap in their bulk band structure, possess conducting edge states protected by symmetries.

Two-dimensional group-14 Xenes consist of a one-atom thick monolayer of X = \ch{Si}, \ch{Ge}, \ch{Sn}, or \ch{Pb} atoms and are named silicene, germanene, stanene and plumbene, respectively. These materials possess a honeycomb crystalline structure akin to graphene, but with the important difference of the presence of a small buckling between X atoms in inequivalent lattice sites \cite{Cahangirov_2009, Xu_2013, Rivero_2014}. Silicene, germanene and stanene exhibit a direct bandgap in their electronic structure, a consequence of the combined action of atomic spin–orbit coupling (SOC) and the crystal field arising from specific orbital hybridizations, while plumbene presents an indirect bandgap between the valence band maximum at the $\Gamma$ point and the conduction band minimum at the $K$ point. In graphene, the atomic SOC is tiny, of the order of $\qty{e-3}{m\electronvolt}$, leading to a negligible bandgap. For heavier group‑14 elements, the atomic SOC values increase: $\qty{3.973}{m\electronvolt}$ for silicene, $\qty{46.3}{m\electronvolt}$ for germanene, $\qty{64.4}{m\electronvolt}$ for stanene \cite{Liu_2011_ham} and $\qty{800}{m\electronvolt}$ for plumbene \cite{Yu_2017}. However, the observable spin splitting --- and thus the topological bandgap --- is not solely determined by the atomic SOC constant; it also depends sensitively on the crystal field induced by the orbital hybridization. This interplay, well documented in the context of Rashba‑like physics \cite{Bihlmayer_2015, MeraAcosta_2020, Bihlmayer_2022}, explains why even modest atomic SOC in silicene and germanene can open a non‑trivial topological gap, while the larger atomic SOC in plumbene yields a trivial gap. Thus, the effective spin‑orbit effects in Xenes are a material‑specific product of atomic SOC and crystal‑field environment, not a simple monotonic trend. The presence of this bandgap in Xenes turns them into semiconductors as opposed to graphene, which is characterised as being a Weyl semimetal.

The aforementioned buckling in Xenes renders the application of an external electric field in a direction perpendicular to the Xene monolayer compelling, since the difference in height between sublattices implies that atoms in the upper sublattice will experience an electric potential opposite in sign to those of the lower sublattice, thus breaking the system's inversion symmetry and allowing bandgap manipulation \cite{Drummond_2012}. On top of this, an extrinsic and an intrinsic Rashba spin-orbit coupling (SOC) occur \cite{Bychkov_1984}, the first as a result of an applied electric field and the latter due to the Xenes' own low-buckled geometries \cite{Liu_2011_ham}.

One of the most distinguished features of silicene, germanene and stanene is that they naturally are topological insulators. In response to a perpendicular electric field, their bandgap can be continuously tuned and closes at a material‑dependent critical electric field value; at which point the system undergoes a topological phase transition (TPT) from a quantum‑spin‑Hall (topological) insulator to a trivial insulator \cite{Ezawa_2012_epj}. This field‑driven switching is of practical interest because it enables reversible control of topological edge conduction --- the operational principle proposed for topological field‑effect transistors, which could provide low‑power, dissipationless switching if materials with sufficiently large gaps and manageable critical fields are found \cite{Ni_2012, Li_2012, Fishman_2020}. Since Xenes in the absence of magnetic fields form systems that preserve time-reversal symmetry (TRS), their topological phases cannot be characterized by the Chern number, as the Berry curvature obeys $\Omega(k) = - \Omega(-k)$, making all quantities involving the integral of $\Omega$ in the whole Brillouin zone vanish \cite{Vanderbilt_2018}. Instead, it is necessary to make use of the $\mathbb{Z}_2$ topological invariant proposed by Kane and Mele \cite{Kane_2005_Z2}. The present work therefore focuses on accurately determining critical fields for silicene, germanene and stanene as a step toward identifying candidates for electrically switchable topological devices.

It is important to note that two‑dimensional layers of group-14 Xenes must be stabilised by a supporting substrate. The interaction with the substrate typically involves charge transfer, hybridisation with the substrate's orbitals, and epitaxial strain, all of which can modify the lattice symmetry, the buckling height, and the resulting band topology \cite{Matusalem_2017, LeLay_2017, Matetskiy_2025, Mihalyuk_2025, Das_2025}. Understanding the topological phase transitions of the pristine, free‑standing Xenes therefore remains a necessary first step, providing a reference limit from which substrate‑driven modifications can be systematically understood. The approach developed in the present work is directly transferable to supported Xenes once a realistic description of the substrate interaction is included, and we envisage such extensions as a natural direction of the present work.

On the experimental side, the synthesis of silicene, germanene, and stanene has been the subject of ongoing investigation. In 2005, Le Lay's group reported the molecular beam epitaxy (MBE) growth of hexagonal \ch{Si} nanoribbons on \ch{Ag}(110), later identified as silicene \cite{Leandri_2005, Davila_2008, Kara_2009, LeLay_2009, Aufray_2010, DePadova_2010}. However, in 2013, Bernard et al. used scanning tunneling microscopy (STM) and grazing incidence X-ray diffraction (GIXD) to show that such growth induces a missing-row reconstruction due to \ch{Ag} atom release \cite{Bernard_2013}. That same year, Colonna et al., after systematically varying STM conditions, suggested that previously reported hexagonal images were tip artifacts \cite{Colonna_2013}. Meanwhile, a 2010 claim of silicene sheets on \ch{Ag}(111) \cite{Lalmi_2010} was met with skepticism: the reported lattice constant of $\qty{1.9 \pm 0.1}{\angstrom}$ was far below density functional theory (DFT) values ($\qty{2.2}{\angstrom}-\qty{2.3}{\angstrom}$) \cite{Cahangirov_2009,Houssa_2010, Cheng_2011, Drummond_2012}, and such high compression ($\sim \qty{17}{\percent}$) was deemed unstable \cite{LeLay_2012}. Moreover, the pristine STM topographies over hundreds of \unit{\square\nm} were later shown to correspond to clean \ch{Ag}(111) with contrast inversion due to tip–sample interactions \cite{Cahangirov_2017, LeLay_2012}.

Reliable silicene synthesis was finally achieved in 2012 on \ch{Ag}(111) \cite{Lin_2012, Vogt_2012, Chiappe_2012, Feng_2012, Chen_2012} and on thin \ch{ZrB2}(0001) films grown by MBE on \ch{Si}(111) \cite{Fleurence_2012}. Subsequent syntheses followed on \ch{Ir}(111) (2013) and on non-metallic \ch{MoS_2} (2014) \cite{Meng_2013, Chiappe_2014}. Notably, Vogt et al. (2012) combined STM, low-energy electron diffraction (LEED), angle-resolved photoemission spectroscopy (ARPES), and DFT-based STM simulations to conclusively demonstrate that the layered \ch{Si} structure qualified as silicene, with measured properties in good agreement with theory \cite{Vogt_2012}.

Germanene and stanene, composed of heavier group-14 atoms (\ch{Ge} and \ch{Sn}), exhibit larger band gaps, making them better candidates for observing the quantum spin Hall effect (QSHE) and topological phase transitions \cite{Lyu_2019, Fuhr_2021}. In 2014, Dávila et al. grew germanene on \ch{Au}(111) by MBE, confirming its structure with STM, LEED, and synchrotron spectroscopy \cite{Dávila_2014}. Furthermore, early reports of germanene growth on \ch{Pt}(111) \cite{Li_2014} have been criticized \cite{Svec_2014}, contending that the observed structures are actually a \ch{Ge3Pt} surface alloy rather than elemental germanene. The following year, the dumbbell-like reconstruction of \ch{Ge} atoms on substrates was demonstrated \cite{Ozcelik_2014}, and in 2015, stanene was also synthesized by MBE \cite{Zhu_2015, Wang_2024}.

The first real-world application of a Xene, in this case silicene, was reported in 2015, when its use in a field-effect transistor (FET) operating at room temperature was demonstrated \cite{Tao_2015}. Beyond conventional electronic applications, non-conventional uses of Xenes have also been explored, such as in spintronics, including spin filters \cite{Tsai_2013, Kharadi_2021} and spin FETs \cite{Wang_2015}\cite{Lyu_2019}. Xene functionalization and doping has also been explored, both theoretically and experimentally, \cite{Zhang_2021, Villarreal_2021, Villarreal_2019, Krawiec_2018}. Furthermore, the exploitation of topological properties in 2D materials and, in particular, topologically protected conducting edge-states, allow for a variety of device applications, such as topological insulator field-effect transistors \cite{Ezawa_2013, Ni_2012, Vandenberghe_2017} and quantum computation \cite{He_2019, Weber_2024}.

In this work, we investigate the topological phase transitions in silicene, germanene, and stanene, focusing on the electric field perpendicular to the layer as the key tuning parameter. This choice is motivated by the fact that an external electric field, applied in a gated device configuration, induces sublattice‑staggered potentials and has been predicted to drive topological phase transitions in group‑IV Xenes. We employ density-functional theory (DFT) in combination with the method of Yu et al. to compute the $\mathbb{Z}_2$ topological invariant, which is particularly well-suited for ab initio approaches since it avoids gauge-fixing ambiguities \cite{Yu_2011}. While previous studies have estimated these critical fields using tight-binding Hamiltonians parameterized from zero-field DFT results \cite{Yu_2018}, such approaches systematically underestimate the transition fields due to their limited treatment of field-induced changes in the electronic structure.

Here, we overcome these limitations by using a fully self-consistent first-principles workflow that performs ab initio simulations for each electric-field strength, followed by Wannierisation and $\mathbb{Z}_2$ invariant evaluation. Firstly, the electric field is introduced directly from first principles rather than from an ad-hoc staggered sublattice term introduced in a tight-binding Hamiltonian, which enables quantitatively predicting critical field values of topological phase transitions using the actual material's dielectric response from both electrons and ions. Secondly, Wannierisation allows extracting a minimal tight-binding Hamiltonian that reproduces the ab initio band structure, enabling efficient computation of topological invariants for arbitrary supercells or different heterostructures. Thirdly, the calculation of $\mathbb{Z}_2$ via the Wannier charge centers (WCC) handles systems without inversion symmetry, unlike parity-based methods (restricted to centrosymmetric crystals) and is not sensitive to gauge ambiguities, unlike direct integration of the Berry connection.

 This sequential approach captures subtle electronic effects that conventional models neglect, yielding significantly more accurate critical field values and establishing a predictive and transferable methodology for exploring topological transitions in two-dimensional materials, paving the way for the rational design of topological devices and quantum technologies. This procedure has already been applied successfully to stanene, yielding $E_z^{\text{cr}} \simeq \qty{0.69}{V/\angstrom}$ \cite{Fuhr_2021}, and this result serves as a benchmark to extend the approach to silicene and germanene.

It is important to note that plumbene behaves qualitatively differently from lighter group-14 Xenes. Band structure orbital character analyses reveal that while the $p_z$ orbital dominates states near the Fermi level in lighter Xenes, plumbene exhibits significant contributions from other orbitals \cite{Yu_2018, Lee_2020}. Because larger interatomic distances in plumbene weaken $s$-orbital bonding, the $s$ antibonding state drops below all $p$ bonding and antibonding states at the $\Gamma$ point. This energy band inversion relative to lighter systems drives plumbene's topologically trivial character. Consequently, to accommodate this altered orbital character near the Fermi level and properly minimize spread during Wannierisation, the basis of maximally localized Wannier functions must be expanded from the one used in this study. Furthermore, an accurate tight-binding representation that faithfully captures its topology requires an expanded orbital basis beyond the Kane-Mele Hamiltonian studied in the present article.

This article is organized as follows. In Section \ref{section:methodology} we present the DFT calculation details (Subsection \ref{section:dft}), the description of the Wannierisation procedure used (Subsection \ref{section:wannierisation}) and the implementation of Yu's method to calculate the $\mathbb{Z}_2$ invariant \cite{Yu_2011} (Subsection \ref{section:Z2}). In Section \ref{section:discussion} a comparison is made between the critical electric field values obtained here and those previously reported, together with those obtained from a theoretical four-band second-nearest-neighbor tight-binding Hamiltonian. Finally, in Section \ref{section:conclusions} the conclusions are outlined.

\section{Methodology}\label{section:methodology}
\subsection{DFT calculations and band structure}\label{section:dft}
The software Quantum-Espresso version 7.3.1 (QE) was used for the DFT calculations \cite{Giannozzi_2009, Giannozzi_2017}. Since we were interested in spin-orbit calculations, full relativistic pseudopotentials were used. In particular, versions 6.3 of the pseudopotentials found in the Pseudopotential library 1.0.0 by Dal Corso \cite{Dal_Corso_2014} for silicon, germanium and tin atoms were used, all of which describe ion-electron interactions in a Kresse-Joubert plane augmented wave method (PAW) \cite{Kresse_1999} and use the Perdew-Burke-Ernzerhof (PBE) exchange-correlation functional for the generalized gradient approximation (GGA) \cite{Perdew_1996}. To simulate a monolayer, the supercell was constructed with $\qty{20}{\angstrom}$ of vacuum above the Xene unit cell to ensure that periodic images of the monolayer did not interact. To exclude any possible influence of van der Waals interactions, test relaxations were performed using the semi-empirical DFT-D3 dispersion correction with Becke-Johnson damping \cite{Grime_2010, Grime_2011}. The resulting optimised geometry was unaltered compared to the plain PBE result, which is expected because the monolayer was modelled in a supercell with $\qty{20}{\angstrom}$ of vacuum separating periodic images, so no interlayer interactions could arise. All calculations presented in the following were therefore performed without explicit dispersion corrections.

Using self-consistent field (SCF) calculations, the kinetic energy cutoff for the wavefunctions was converged to $\qty{60}{Ry}$, while the kinetic energy cutoff for the charge density and potential was set to an eightfold multiplicity of the previous result, that is $\qty{480}{Ry}$. Succesful convergence of the total energy with respect to the number of k-points was attained for a $13 \times 13 \times 1$ k-point grid. Regarding the structural optimization, the total energy was minimised with respect to variations of the lattice parameter $a$ and buckling $\ell$, resulting in a $D_{3d}$-symmetric structure, characterised by the values shown in table \ref{tab:dftvalues}, which are in good agreement with the structural first-principles values obtained in \cite{Liu_2011_ham}. The in-plane lattice constant $a$ increases systematically from silicene to stanene, consistent with the larger covalent radius of the heavier group‑14 elements. The buckling height $\ell$ follows the same trend, reflecting the increasing tendency of the heavier atoms to favour $sp^3$-like hybridization over purely planar $sp^2$ bonding; this effect is a direct consequence of the reduced energy splitting between the valence $s$ and $p$ orbitals as one descends the group.

\ctable[
caption = {Lattice parameter $a$, buckling $\ell$ and bandgap $E_g$ obtained in this work for each material, in good accordance with those previously reported in \cite{Liu_2011_ham}.},
label = tab:dftvalues,
pos = hptb
]{@{}lcccccc@{}}{
}{                                                                \FL
Materials & $a (\unit{\angstrom})$ & $\ell (\unit{\angstrom})$ & $E_g (\unit{meV})$ & $a\textsuperscript{ref} (\unit{\angstrom})$ & $\ell\textsuperscript{ref} (\unit{\angstrom})$ & $E_g\textsuperscript{ref} (\unit{meV})$  \ML
Silicene  & $3.867$ & $0.227$ & $1.5$  & $3.86$ & $0.23$ & $1.55$  \NN
Germanene & $4.043$ & $0.340$ & $24.2$ & $4.02$ & $0.33$ & $23.9$  \NN
Stanene   & $4.675$ & $0.428$ & $77.3$ & $4.70$ & $0.40$ & $73.5$  \LL
}

\begin{figure}[thpb]
    \centering
    \begin{subfigure}[t]{0.65\textwidth}
        \centering
        \phantomcaption
        \stackinset{l}{4.2mm}{t}{3.5mm}{\captiontext*{}}
        {
        \includegraphics[width=\textwidth]{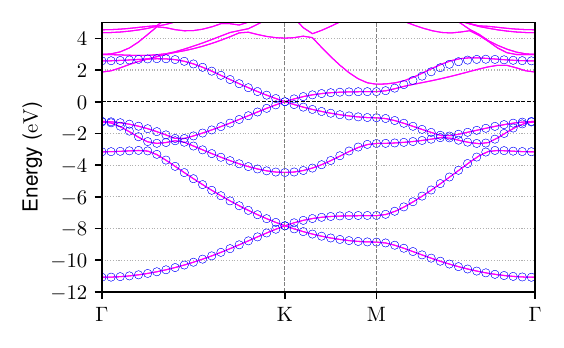}
        }
    \end{subfigure}
    \begin{subfigure}[t]{0.3\textwidth}
        \centering
        \raisebox{0.8cm}{\includegraphics[width=\textwidth]{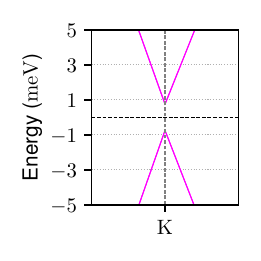}}
    \end{subfigure}
    \par
    \begin{subfigure}[t]{0.65\textwidth}
        \centering
        \phantomcaption
        \stackinset{l}{4.2mm}{t}{3.5mm}{\captiontext*{}}
        {
        \includegraphics[width=\textwidth]{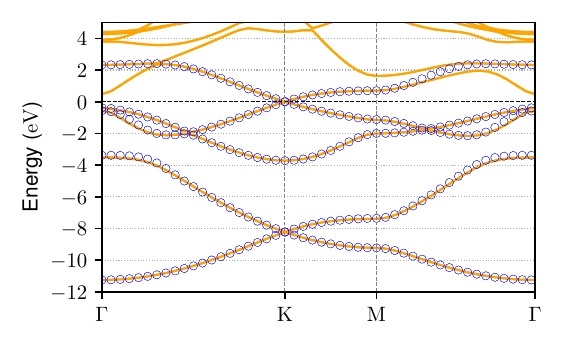}
        }
    \end{subfigure}
    \begin{subfigure}[t]{0.3\textwidth}
        \centering
        \raisebox{0.8cm}{\includegraphics[width=\textwidth]{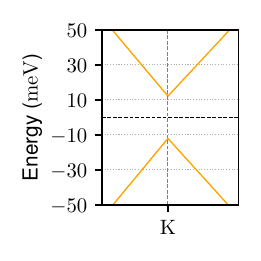}}
    \end{subfigure}
    \par
    \begin{subfigure}[t]{0.65\textwidth}
        \centering
        \phantomcaption
        \stackinset{l}{4.2mm}{t}{3.5mm}{\captiontext*{}}
        {
        \includegraphics[width=\textwidth]{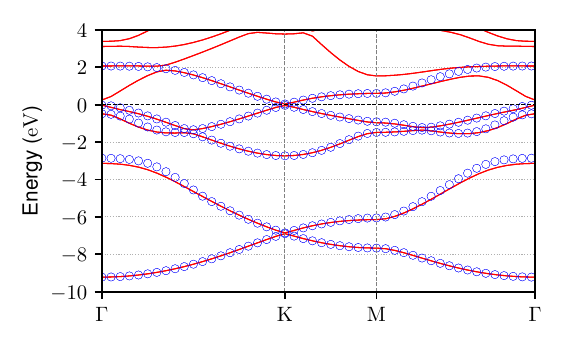}
        }
    \end{subfigure}
    \begin{subfigure}[t]{0.3\textwidth}
        \centering
        \raisebox{0.8cm}{\includegraphics[width=\textwidth]{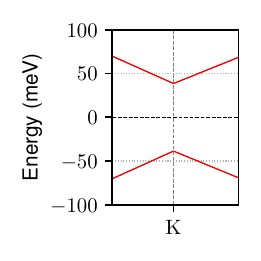}}
    \end{subfigure}
       \caption{Left: Comparison of DFT‑calculated band structures for (a) silicene, (b) germanene and (c) stanene with the Wannier‑interpolated band structures (blue circles). Right: Zoomed-in view of the direct bandgap at the $\mathrm{K}$ high-symmetry point.}
    \label{fig:band_structures}
\end{figure}

In order to have an accurate description of the band structure, a band calculation with fixed occupations was performed using the $\Gamma-\mathrm{K}-\mathrm{M}-\Gamma$ path in the first Brillouin zone. The calculated band structures for silicene, germanene and stanene are presented in figure \ref{fig:band_structures}, each including a corresponding zoomed-in view of the bandgap region. Without spin–orbit coupling, the low‑buckled honeycomb lattice would host Dirac‑like cones at the high-symmetry K point, similar to graphene. The inclusion of SOC, captured by our fully relativistic DFT calculations, opens a substantial gap at K whose magnitude grows rapidly from silicene to stanene. The size of the bandgap obtained was $\qty{1.5}{meV}$, $\qty{24.2}{meV}$ and $\qty{77.3}{meV}$ for silicene, germanene and stanene, respectively. The 20 lower $d$-type bands for germanene and stanene are not shown since they do not affect the physics at the Fermi level. In table \ref{tab:dftvalues}, these values are compared with those reported in \cite{Liu_2011_ham}, where clear physical trends emerge from the data. Taken together, these results confirm that our structural and electronic description of free‑standing Xenes faithfully reproduces the established intrinsic properties, and provides a clean baseline for the topological analysis that follows.

The direct bandgap at the high-symmetry point K in Xenes is a particular focus of our work, because a closing of the bandgap is a necessary condition for a topological phase transition in which the symmetries of the system are preserved throughout the whole process \cite{Ezawa_2013, Bernevig_2013}. For this reason, the variation of this bandgap with the intensity of the applied perpendicular electric field was analysed for all three materials directly at the K-point.

To simulate an electric field in self-consistent field (SCF) and non-self-consistent field (NSCF) calculations a saw-like potential was implemented, taking into account the dipole correction proposed by Bengtsson \cite{Bengtsson_1999}. The potential was carefully placed so that the zone where the potential abruptly decreased was exactly at the midpoint between the mirror images of the Xene being studied. Electric field amplitudes were varied up to $E_z = \qty{0.8}{V/\angstrom}$. Following established literature \cite{Ezawa_2012_epj, Houssa_2016}, the evolution of the bandgap $E_g$ is symmetric in field polarity for freestanding group‑14 Xenes $E_g(E_z) = E_g(-E_z)$, which is why only positive $E_z$ values are shown.

\begin{figure}[thpb]
    \centering
    \begin{subfigure}[t]{3.6in}
        \centering
        \includegraphics[scale=1]{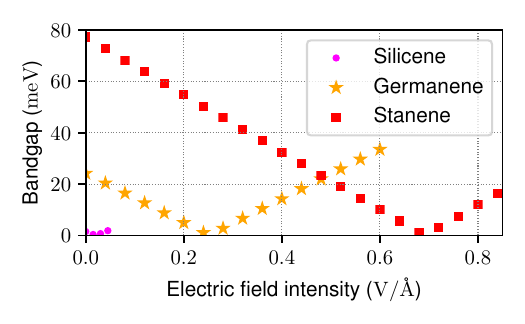}
        \subcaption{}
        \label{fig:all_bandgaps}
    \end{subfigure}
    \hfill
    \begin{subfigure}[t]{2in}
        \centering
        \raisebox{0cm}{\includegraphics[scale=1]{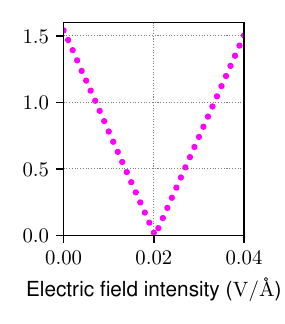}}
        \subcaption{}
        \label{fig:silicene_bandgap}
    \end{subfigure}
       \caption{Bandgap variation with the intensity of the perpendicularly applied electric field in silicene, germanene and stanene \subref{fig:all_bandgaps} and zoomed-in view for silicene \subref{fig:silicene_bandgap}.}
    \label{fig:all_bandgaps_QE}
\end{figure}

It was seen that contact between the valence and conduction bands occured at $\qty{0.020}{V/\angstrom}$, $\qty{0.250}{V/\angstrom}$ and $\qty{0.690}{V/\angstrom}$, for silicene, germanene and stanene, respectively, which is shown in figure \ref{fig:all_bandgaps_QE}. The observed gap variation under an out-of-plane electric field can be understood as a competition between two symmetry-inequivalent gap-tuning mechanisms. In the absence of an external field, the intrinsic spin-orbit coupling opens a non‑trivial gap at the $K$ point. Applying a perpendicular electric field breaks the in‑plane inversion symmetry of the buckled lattice and introduces a staggered sublattice potential. As the field strength increases, the two contributions progressively cancel, driving the system towards a band‑gap closing at a critical field $E_z^{\text{cr}}$. At this point the system is expected to undergo a topological phase transition (TPT) between the quantum spin Hall phase and a trivial insulating phase. The critical field therefore scales with the intrinsic SOC strength: it is smallest for silicene, whose SOC gap is only $\qty{1.5}{meV}$ ($\qty{0.020}{V/\angstrom}$), and largest for stanene, where the SOC gap reaches $\qty{77.3}{meV}$ ($\qty{0.690}{V/\angstrom}$). The value obtained for stanene coincides with the transition field reported previously using the same methodology \cite{Fuhr_2021}, and the same physical mechanism allows us to predict the previously unreported critical fields for silicene and germanene.

\subsection{Wannierisation}\label{section:wannierisation}
In order to obtain a tight-binding representation, suitable for later calculation of the topological invariant, it is desirable to work in a representation with orthogonal basis wavefunctions that are spatially localized. This is where Wannier functions $\ket{\alpha i}$ come in, where $\alpha$ is the function index and $i$ denotes the position $\mathbf{R}_i$ of the cell at which the function is located. These Wannier functions are defined as inverse Fourier transforms of a unitary mixing of a set of $J$ Bloch eigenstates $\ket{\psi_{m \mathbf{k}}}$ in the Brillouin zone, that is
\begin{equation}
    \ket{\alpha i}
    =
    \frac{V}{(2\pi)^3}
    \int_{\mathrm{BZ}}
    \left[
        \sum_{m=1}^{J} U^{(\mathbf{k})}_{m \alpha} \ket{ \psi_{m \mathbf{k}} }
    \right]
    \mathrm{e}^{-\mathrm{i}\mathbf{k}.\mathbf{R}} \, \mathrm{d}\mathbf{k}\,,
\end{equation}
where $U^{(\mathbf{k})}$ is a unitary matrix that mixes the Bloch states at a given $\mathbf{k}$ \cite{Pizzi_2020}. The matrix $U^{(\mathbf{k})}$ is not unique and a convenient choice, as proposed by Marzari and Vanderbilt \cite{Marzari_1997}, is to select one that minimizes the sum $\Omega_{\text{tot}}$ of the spatial spreads of the $J$ Wannier functions, defined as
\begin{equation}\label{eq:spread}
\Omega_{\text{tot}}
=
\sum_{\alpha=1}^J
\left[
    \braketOP
        {\alpha \mathbf{0}}
        {\mathbf{r} \cdot \mathbf{r}}
        {\alpha \mathbf{0}}
    -
    \abs{\braketOP
        {\alpha \mathbf{0}}
        {\mathbf{r}}
        {\alpha \mathbf{0}}
        }^2
\right]\,.
\end{equation}

The spread can be decomposed into two terms $\Omega_{\text{tot}} = \Omega_I + \tilde{\Omega}$, a gauge-invariant term
\begin{equation}\label{eq:invariant_spread}
\Omega_I
=
\sum_{\alpha}
\left[
    \braketOP
        {\alpha \mathbf{0}}
        {\mathbf{r} \cdot \mathbf{r}}
        {\alpha \mathbf{0}}
    -
    \sum_{\beta i} \abs{ \braketOP
        {\beta i}
        {\mathbf{r}}
        {\alpha \mathbf{0}}}^2
\right]
\end{equation}
and a gauge-dependent term
\begin{equation}\label{eq:gauge_dependent_spread}
\tilde{\Omega}
=
\sum_{\alpha}
\sum_{\beta i \neq \alpha \mathbf{0}}
\abs{ \braketOP
        {\beta i}
        {\mathbf{r}}
        {\alpha \mathbf{0}}
    }^2
\end{equation}
which can be minimised by varying the $U^{(\mathbf{k})}$ matrices, thereby obtaining maximally localized Wannier functions (MLWFs).

To obtain these MLWFs, a noncollinear spin-polarized SCF calculation with spin-orbit coupling was performed, using a $13 \times 13 \times 1$ k-point grid. This was followed by a NSCF calculation using 441 k-points ($21 \times 21 \times 1$). At this point, the software Wannier90 version 3.1.0 (W90) \cite{Pizzi_2020} was used in order to obtain maximally-localized Wannier functions (MLWFs), following the method outlined by Souza, Marzari and Vanderbilt for entangled energy bands \cite{Souza_2001}.

The post-processing interface pw2wannier90 between QE and W90 was used to construct the matrix elements $M_{mn}^{(\mathbf{k},\mathbf{b})} = \braket{m,\mathbf{k}}{n,\mathbf{k}+\textbf{b}}$ of initial overlaps between the periodic parts of the Bloch states $\ket{m,\mathbf{k}}$ and $\ket{n,\mathbf{k}}$ of neighboring k-points, as well as the matrix elements $A_{mn}^{(\mathbf{k})} = \braket{\psi_{m\mathbf{k}}}{\varphi_{n}}$ of the projections of the Bloch states $\ket{\psi_{m\mathbf{k}}}$ onto the initial orbitals $\ket{\varphi_n}$ \cite{Pizzi_2020}.

Subsequently, the main W90 routine was used to disentangle the bands and find the MLWFs, minimizing the spread $\Omega$. The top of the energy window for the disentanglement procedure was set to $\qty{3}{\electronvolt}$ above the highest occuppied state in the case of silicene and germanene, and $\qty{2.5}{\electronvolt}$ in the case of stanene. For germanene and stanene, it was also important to set the bottom of the energy window just below the 8 bonding energy bands closest to the Fermi level, since it is not our interest to represent the lower $d$-character bands.

For each spin, three $s$ orbitals centered on the bonds between first-neighbors and two $p_{z}$ orbitals centered on each of the atoms of the unit cell were used as initial guesses of the wanierisation procedure, in a manner similar to that used by Jung and MacDonald for graphene \cite{Jung_2013}, thus obtaining 10 spin-polarized MLWFs for each material.

\begin{figure}[thpb]
    \centering
    \begin{subfigure}[t]{\textwidth}
        \centering
        \includegraphics[width=0.36\textwidth]{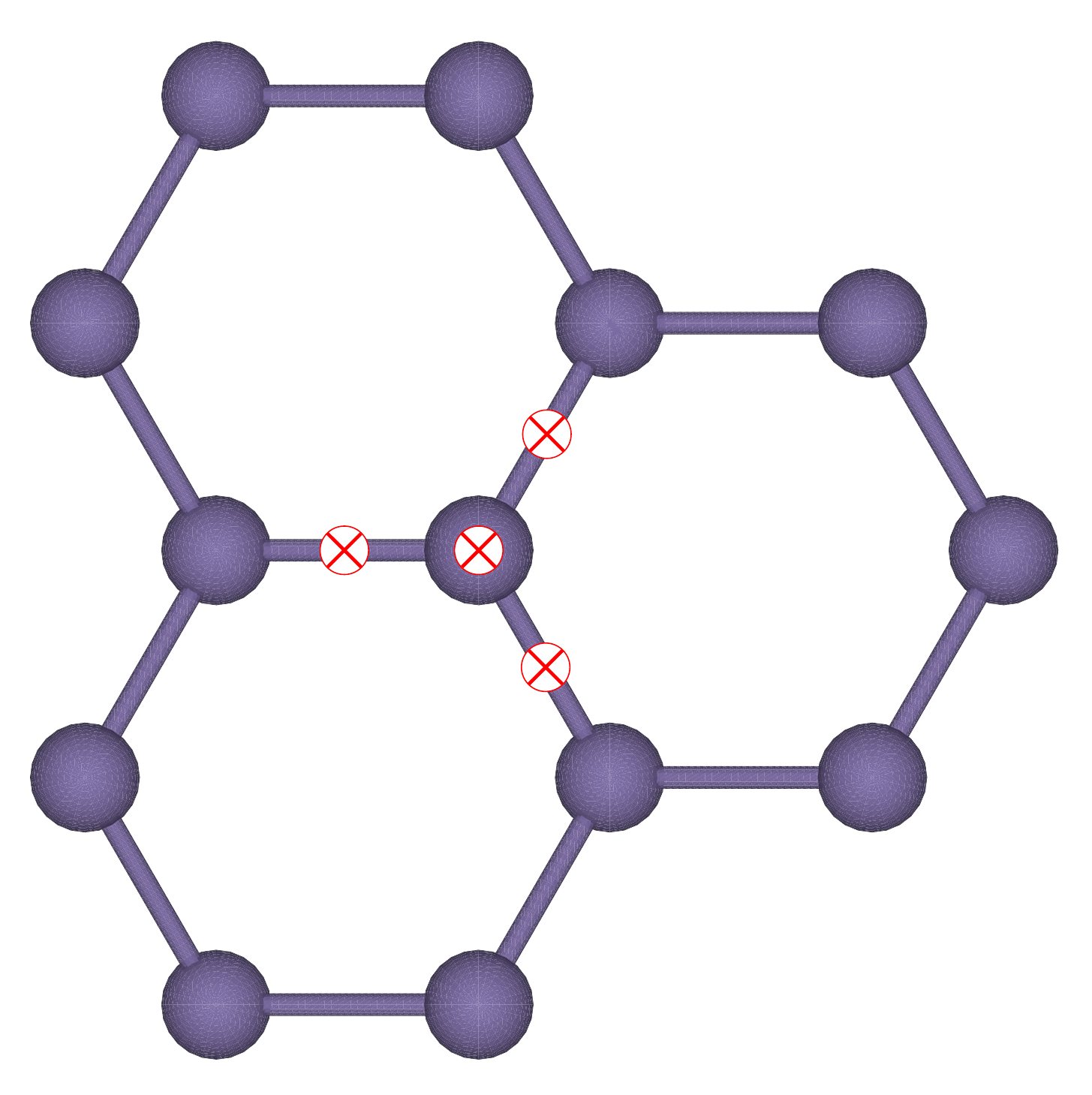}
        \caption{}
        \label{fig:centers_above}
    \end{subfigure}
    \par\bigskip
    \begin{subfigure}[t]{\textwidth}
        \centering
        \includegraphics[width=0.36\textwidth]{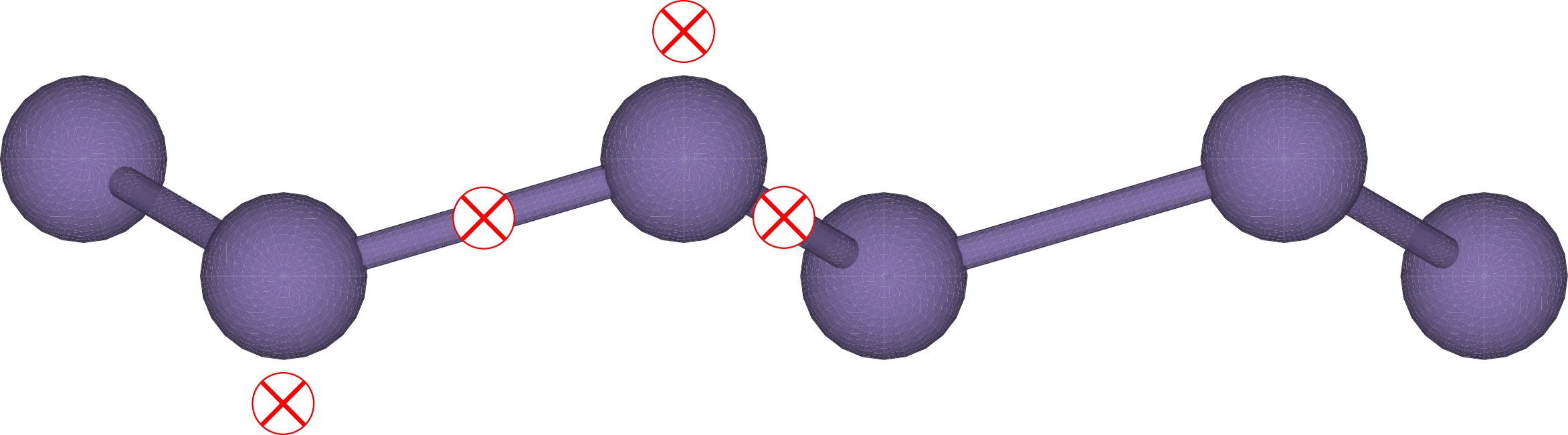}
        \caption{}
        \label{fig:centers_side}
    \end{subfigure}
       \caption{Germanene Wannier centers, marked with red crosses, seen from above \subref{fig:centers_above} and from the side \subref{fig:centers_side}. The figures were obtained using VESTA \cite{Momma_2011}.}
       \label{fig:germanene_centers}
\end{figure}

\begin{figure}[thpb]
    \centering
     \begin{subfigure}{0.45\linewidth}
        \centering
        \includegraphics[width=0.8\linewidth]{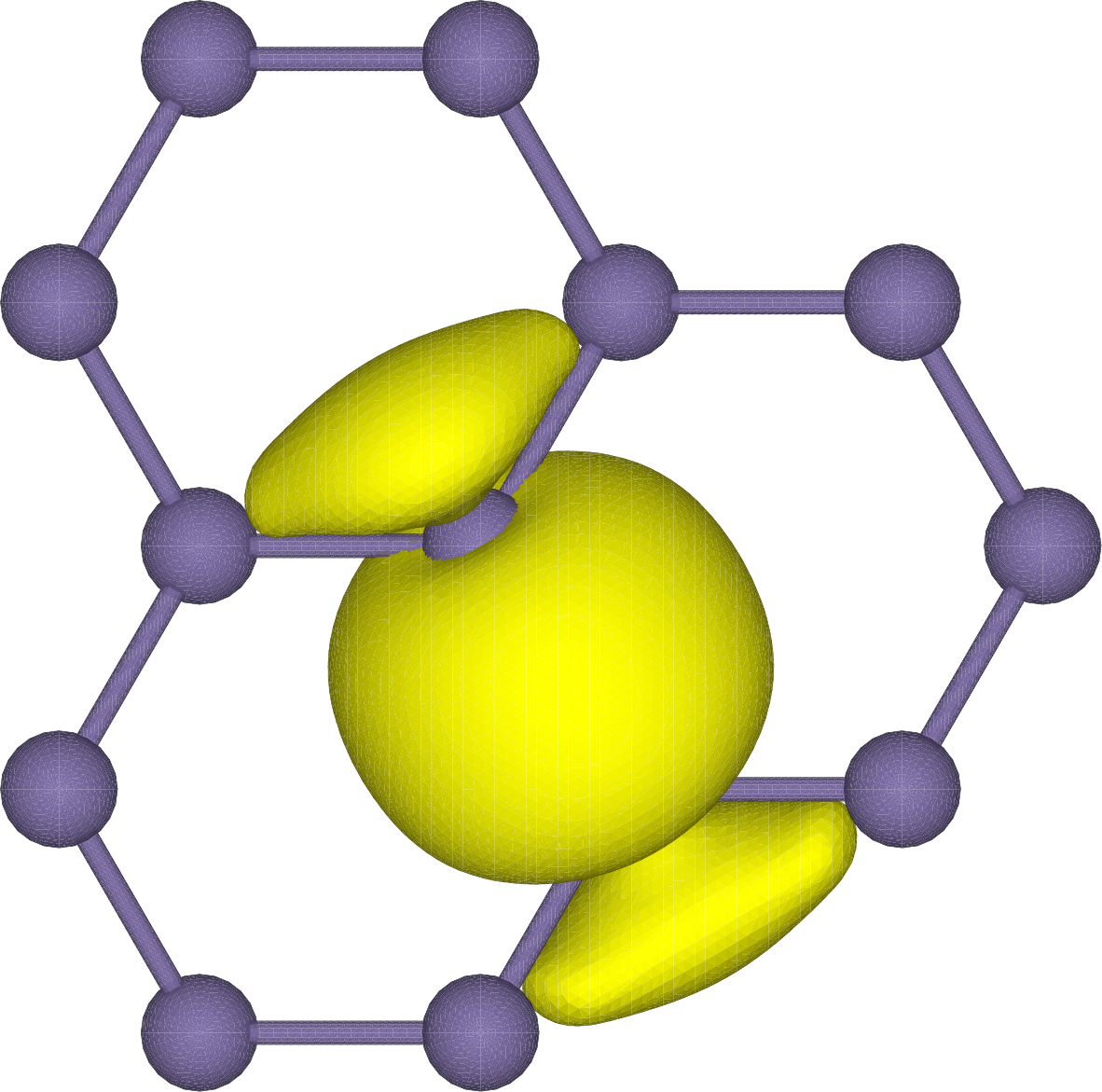}
        \subcaption{}
        \label{fig:orbital1_top}
    \end{subfigure}
    \hfill
    \begin{subfigure}{0.45\linewidth}
        \centering
        \includegraphics[width=0.8\linewidth]{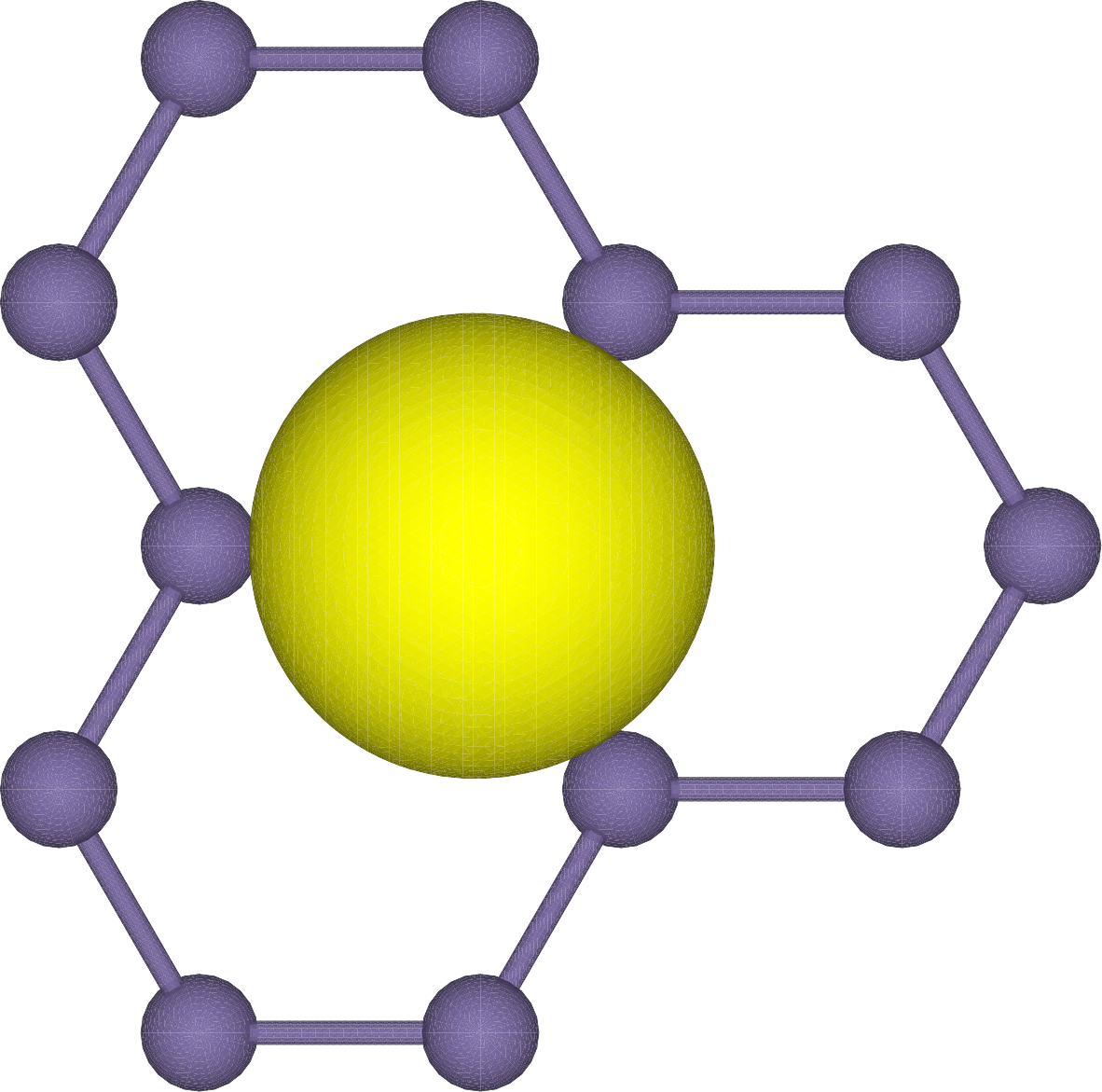}
        \subcaption{}
        \label{fig:orbital2_top}
    \end{subfigure}
    \par\bigskip
    \begin{subfigure}{0.45\linewidth}
        \centering
        \includegraphics[width=0.8\linewidth]{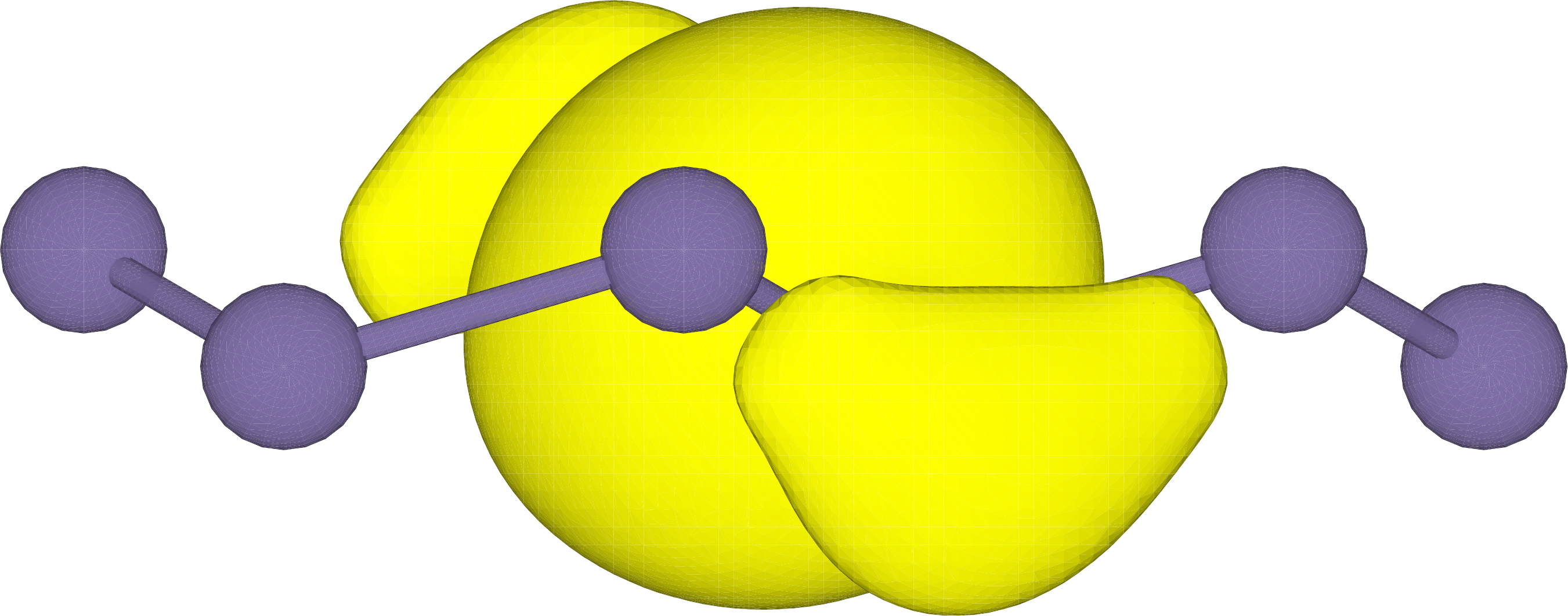}
        \subcaption{}
        \label{fig:orbital1_side}
    \end{subfigure}
    \hfill
    \begin{subfigure}{0.45\linewidth}
        \centering
        \includegraphics[width=0.8\linewidth]{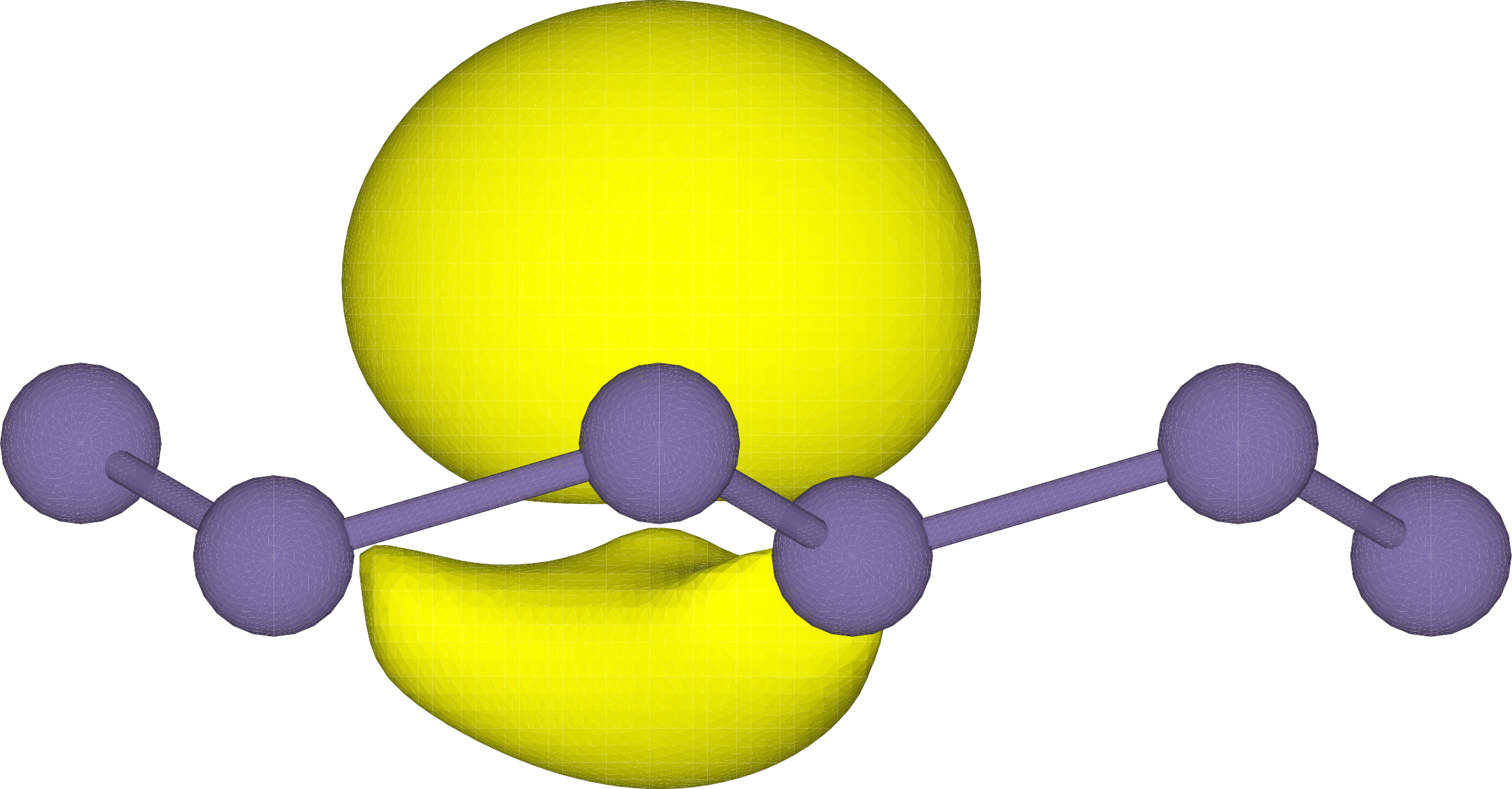}
        \subcaption{}
        \label{fig:orbital2_side}
    \end{subfigure}
    \caption{Germanene bond-centered (left) and atom-centered (right) maximally localized Wannier orbitals seen from above (top) and from the side (bottom). The bond-centered MLWFs are akin to $\sigma$ bonds in graphene while the atom-centered MLWFs bear resemblance to the $p_z$ orbitals in graphene, responsible for $\pi$-bonding. The figures were obtained using VESTA \cite{Momma_2011}.}\label{fig:orbitals}
\end{figure}

After the Wannierisation procedure carried out with the software Wannier90, the resulting Wannier orbitals for silicene, germanene and stanene proved to be nearly identical, which is why only the isosurface plots of the converged Wannier orbitals in the case of germanene are shown in figure \ref{fig:orbitals}. As can be seen in figure \ref{fig:germanene_centers}, six of these MLWFs are centered exactly at the midpoint between neighboring atoms, while the remaining four MLWFs are vertically displaced with respect to the Xene atoms. The magnitude of this displacement is $\qty{0.601}{\angstrom}$, $\qty{0.644}{\angstrom}$ and $\qty{0.711}{\angstrom}$ in the cases of silicene, germanene and stanene, respectively. The shapes of the 6 bond-centered and 4 atom-centered MLWFs, shown in figure \ref{fig:orbitals}, closely resemble $\sigma$ bonds and $p_z$ orbitals, respectively, but with a small asymmetry along the z-direction due to the buckling present in Xenes, compared with pure $\sigma$ and $p_z$ MLWFs in the case of graphene; cf. figure 6 in \cite{Marzari_2012}.

The converged total spread of the 10 spin-polarized MLWFs for each material is detailed in table \ref{tab:spreads}. It can be seen that the total spread $\Omega_{\text{tot}}$ obtained was larger for stanene than for both germanene and silicene. This is mainly related to the increase in the invariant spread $\Omega_{I}$ that occurs as a result of the smaller energy range spanned by the 10 relevant bands in the case of stanene. Furthermore, a comparison betwen the band structures calculated using DFT (solid lines) and the Wannier‑interpolated bands (blue circles) are shown in figure \ref{fig:band_structures}. For all three materials the Wannier‑interpolated bands faithfully reproduce the DFT eigenvalues, especially in the crucial low‑energy region near the $K$ point, confirming that the Wannier basis obtained can be safely used for the topological invariant calculations that follow.

W90 software produces hopping coefficients in real space between localized Wannier orbitals. Thus, by Fourier transforming into reciprocal space, a Wannier tight-binding Hamiltonian (WTBH) can be obtained. This Hamiltonian was later used to obtain the $\mathbb{Z}_2$ topological invariant using the method described by Yu et al. \cite{Yu_2011}.

\ctable[
cap = {Spreads obtained in the Wannierisation procedure for each Xene.},
caption = {Spreads obtained in the Wannierisation procedure for each Xene. $\Omega_{\sigma}$ and $\Omega_{p_z}$ represent the mean spread of each MLWF of the given type, while $\Omega_{\text{tot}}$ and $\Omega_{I}$ represent the total and invariant spread, respectively.},
label = tab:spreads,
pos = hptb
]{@{}lcccc@{}}{
}{                                                                \FL
Materials & $\Omega_{\sigma} (\unit{\angstrom^2})$ & $\Omega_{p_z} (\unit{\angstrom^2})$ & $\Omega_{\text{tot}} (\unit{\angstrom^2})$ & $\Omega_{I} (\unit{\angstrom^2})$  \ML
Silicene  & $1.93$ & $3.19$ & $24.33$ & $21.91$  \NN
Germanene & $2.29$ & $2.63$ & $24.28$ & $22.57$   \NN
Stanene   & $2.92$ & $2.98$ & $29.44$ & $27.38$   \LL
}

\subsection{Calculation of the $\mathbb{Z}_2$ invariant}\label{section:Z2}
Since the systems considered preserve time-reversal symmetry but break inversion symmetry, we employed the numerical method proposed by Yu et al. to compute the $\mathbb{Z}_2$ invariant \cite{Yu_2011}. This approach offers the key advantage of avoiding gauge-fixing requirements, unlike alternative methods that depend on fixing the gauge of the eigenfunctions in half of the Brillouin zone to calculate the Berry connection \cite{Fukui_2007}. Instead, this method hinges on tracking the evolution of the Wannier function centers in the subspace of occupied states for effective one-dimensional systems of fixed crystal momentum $k_y$.

Let the Hamiltonian of a crystal system with time reversal and translational symmetry expressed in a Wannier function basis $\{\ket{\alpha i}\}$ be
\begin{equation}\label{eq:hamiltonian}
    H = \sum_{\alpha \beta} \sum_{i j} h^{\alpha \beta}_{ij} \ket{\alpha i} \bra{\beta j} + \text{H.c.}\,,
\end{equation}
where $\alpha$ is the index of the Wannier function and $i$ denotes the fact that the function is located at the $i$-th lattice site position $\mathbf{R}_i$.

Its Bloch eigenstates $\ket{\psi_{n\mathbf{k}}}$ can be written in terms of the Fourier transforms $\ket{\alpha \mathbf{k}}$ of the Wannier basis
\begin{equation}\label{eq:bloch_eigenstates}
    \ket{\psi_{n \mathbf{k}}} = \sum_{\alpha} g_{n \alpha}(\mathbf{k}) \ket{\alpha \mathbf{k}}\,,
\end{equation}
where $g_{n\alpha}(\mathbf{k})$ are the coefficients of the expansion and $\ket{\alpha \mathbf{k}} = \frac{1}{\sqrt{N_{cell}}} \sum_i \mathrm{e}^{i \mathbf{k}\cdot\mathbf{R}_i} \ket{\alpha i}$.

Since we have a 2D system, each wave vector $k_y$ defines an effective 1D system with a periodic condition. Let $\hat{X}$ be the position operator for this 1D system defined by the following equation
\begin{equation}\label{eq:position_operator}
    \hat{X}(k_y) = \sum_{i \alpha} \mathrm{e}^{-i \delta k_x R_{i,x}} \ket{\alpha i} \bra{\alpha i}\,,
\end{equation}
where $\delta k_x = \frac{2\pi}{N_x a_x}$, $N_x$ is the number of unit cells, $a_x$ is the lattice constant along the $x$ direction and $R_{i,x}$ labels the unit cell.

The position operator projected onto the subspace of occupied states is 
\begin{equation}\label{eq:projected_position_operator}
    \hat{X}_{P}(k_y)
    = \hat{P}_{k_y} \hat{X} \hat{P}_{k_y}\,,
\end{equation}
with $\hat{P}_{k_{y}}$ the projection operator for the occupied subspace
\begin{equation}\label{eq:projection_operator}
    \hat{P}_{k_{y}} = \sum_{n \in o, k_x} \ket{\psi_{n \mathbf{k}}} \bra{\psi_{n \mathbf{k}}}\,,
\end{equation}
where $o$ denotes the $2N$ occupied bands.

The MLWFs of the occupied subspace can be considered as eigenstates of the projected position operator $\hat{X}_{P}$, with eigenvalues given by the centers of the MLWFs \cite{Kivelson_1982}. Thus, to track the Wannier function centers we want to obtain the eigenvalues of the operator $\hat{X}_{P}$ which, substituting Eqs. \eqref{eq:position_operator}, \eqref{eq:projection_operator} and \eqref{eq:bloch_eigenstates} into Eq. \eqref{eq:projected_position_operator}, is given by
\begin{equation}
    \hat{X}_{P}(k_y)
    = \sum_{k_x k_x'} \delta(k_x + \delta k_x - k_x') \sum_{nm \in o} \ket{\psi_{n k_x k_y}}\bra{\psi_{m k_x' k_y}} \sum_{\alpha} g_{n\alpha}^{*}(k_x) g_{m\alpha}(k_x')\,,
\end{equation}
which can also be expressed in matrix form as
\begin{equation}
    \hat{X}_{P}(k_y) =
    \begin{bmatrix}
        0           & F_{0,1}   & 0         & 0         & 0     & 0\\
        0           & 0         & F_{1,2}   & 0         & 0     & 0\\
        0           & 0         & 0         & F_{2,3}   & 0     & 0\\
        0           & 0         & 0         & 0         & \cdots & 0\\
        0           & 0         & 0         & 0         & 0     & F_{N_x-2,N_x-1}\\
        F_{N_x-1,0} & 0         & 0         & 0         & 0     & 0
    \end{bmatrix}\,,
\end{equation}
where
\begin{equation}
    F_{i,i+1}^{nm}(k_y) = \sum_{\alpha} g_{n\alpha}^{*}(k_{x,i},k_y) g_{m\alpha}(k_{x,i+1},k_y)
\end{equation}
are the elements of the $2N \times 2N$ $F_{i,i+1}$ matrices and $k_{x,i} = \frac{2 \pi i}{N_x a_x}$.

This way, it can be seen that $\hat{X}_{P}$ acts as a cyclic shift modulated by the matrices $F_{i,i+1}$ with respect to the discrete set of $N_x$ $k_{x,i}$-points, with periodic boundary condition $k_{x,N_x} \equiv k_{x,0}$. Thus, defining the product of all the $F_{i,i+1}$ matrices as the $2N \times 2N$ matrix
\begin{equation}\label{eq:D_matrix}
    D(k_y) = \prod_{i=0}^{N_x-1} F_{i,i+1}(k_y)\,,
\end{equation}
it can be proven that the eigenvalues $\lambda_{n,j}^{P}$ of $\hat{X}_P$ are related to the $2N$ eigenvalues $\lambda_n^D$ of the $D$ matrix by the equation
\begin{equation}\label{eq:P_eigenvalues}
    \lambda_{n,j}^{P} = \sqrt[\leftroot{-2}\uproot{2}N_x]{\lambda_n^D} = \sqrt[\leftroot{-2}\uproot{2}N_x]{\abs{\lambda_n^D}} \mathrm{e}^{i(\theta_n^D + 2\pi j)/N_x} = \mathrm{e}^{i(\theta_n^D + 2\pi j)/N_x}\,,
\end{equation}
where $j=1,2,\dots, N_x$. In the last step we have used that $\abs{\lambda_n^D}=1$ since $D$ is a unitary matrix.

It can also be proved that the matrix elements $F_{i, i+1}^{nm}$ can be written as an inner product, namely
\begin{equation}
    F_{i,i+1}^{nm}(k_y) = \braket{n, k_{x,i}, k_y}{m, k_{x,i+1}, k_y}\,,
\end{equation}
where $\ket{n, \mathbf{k}} = \mathrm{e}^{-i \mathbf{k} \cdot \mathbf{r}} \ket{\psi_{n,\mathbf{k}}}$ are the periodic parts of the Bloch functions.

For infinitesimal $\delta_k$, $F_{i, i+1} = \mathrm{e}^{-i A_{i, i+1} \delta k}$, where $A_{i, i+1}^{nm}$ is the non-Abelian Berry connection $A_{i, i+1}^{nm} = i \frac{\bra{n,k_{x,i},k_y} (\ket{m,k_{x,i+1},k_y} - \ket{m,k_{x,i},k_y})}{\delta k}$. Hence, in the continuum limit $\delta_k \to 0$, the ordered product of infinitesimal evolutions ---Equation \eqref{eq:D_matrix}--- becomes a path-ordered exponential
\begin{equation}\label{eq:Wilson_loop}
    D(k_y) = \prod_{i=0}^{N_x-1} F_{i,i+1} = \prod_{i=0}^{N_x-1} \mathrm{e}^{-i A_{i,i+1} \delta k} \underset{\delta k \to 0}{=} P \exp{ \left[ \int_{\mathcal{C}_{k_y}} -i A(k_x) \,\mathrm{d}k_x \right] }\,,
\end{equation}
where $P$ is the ordering operator.

The right hand side of Equation \eqref{eq:Wilson_loop} is the $U(2N)$ Wilson loop of the non-Abelian Berry connection $A(k)$ along a contour $\mathcal{C}_{k_y}$ at fixed $k_y$ that goes across the Brillouin zone from $k_x=-\pi$ to $k_x=\pi$. Thus, Yu et al. demonstrated that the Wilson loop can be numerically calculated using the equivalent discrete expression given by equation \eqref{eq:D_matrix} directly from the wavefunctions, without resorting to fixing the gauge in the Brillouin zone. Moreover, equation \eqref{eq:P_eigenvalues} implies that the evolution of the $2N$ Wannier centers of the occupied subspace with $k_y$ can be described by the $2N$ phases $\theta_{n}^{D}$ of the eigenvalues $\lambda_{n}^{D}$ of the $D$ matrix
\begin{equation}
    \theta_{n}^{D}(k_y) = \Imag\left[{\log{\lambda_{n}^{D}(k_y)}}\right]\,.
\end{equation}

Considering a 2D parameter space formed by the crystal momentum $k_y \in [0,\pi]$ in the horizontal axis and the phase value $\theta \in [-\pi,\pi]$ in the vertical axis, where $\theta = - \pi$ is equivalent to $\theta = \pi$, we can define a cylinder. Because of time-reversal symmetry, the phases $\theta_{n}^{D}$ of the eigevalues $\lambda_{n}^{D}$ must form Kramer pairs at the time-reversal invariant momentum values $k_y=0$ and $k_y=\pi$. Hence, the evolution of a pair of phases with $k_y$ must circle around the cylinder an integer amount of times. It is important to note that every crossing between the two phases that comprises a pair consitutes a degeneracy that is not protected by time-reversal symmetry, which can be transformed into an anticrossing by a continuous transformation of the Hamiltonian. This implies, for instance, that the case in which the winding number of a given pair is two is topologically equivalent to a case in which the winding number is zero. This means that the topological invariant that characterizes the system is a $\mathbb{Z}_2$ quantity $\Delta$, which is simply given by
\begin{equation}\label{eq:invariant}
    \Delta = \sum_{n=1}^{2N} M_n \bmod 2\,,
\end{equation}
where $M_n$ is the number of times each phase $\theta_{n}^{D}$ crosses a given reference horizontal line $\theta=\theta_0$ in its evolution from $k_y=0$ to $k_y=\pi$.

The algorithm explained above was used to calculate via a custom Python script the phases $\theta_{n}^{D}$ of the eigenvalues $\lambda_{n}^{D}$ of the $D$ matrix for different values of $k_y$ between $0$ and $\pi$, thus obtaining a plot of $\theta_{n}^{D}$ vs $k_y$, and this calculation was repeated for a range of values of perpedicularly applied electric field in all the three different materials considered.

\section{Results and discussion}\label{section:discussion}

\begin{figure}[thpb]
    \centering
     \begin{subfigure}{0.49\linewidth}
        \centering
        \phantomcaption
        \stackinset{l}{3mm}{t}{3mm}{\captiontext*{}}{\includegraphics[width=\linewidth]{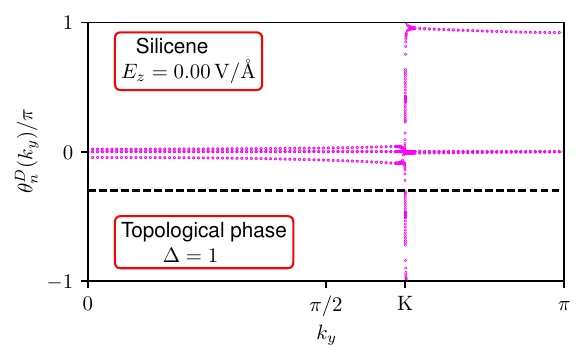}}
        \label{fig:silicene_topo}
    \end{subfigure}
    \hfill
    \begin{subfigure}{0.49\linewidth}
        \centering
        \phantomcaption
        \stackinset{l}{3mm}{t}{3mm}{\captiontext*{}}{\includegraphics[width=\linewidth]{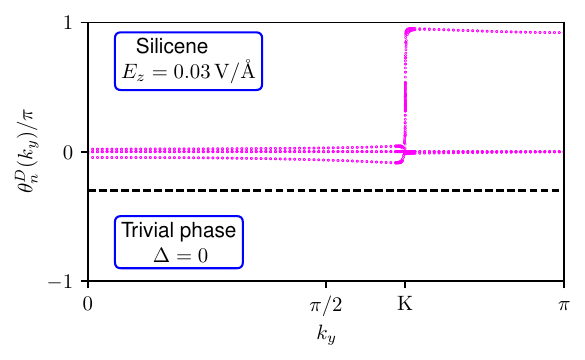}}
        \label{fig:silicene_triv}
    \end{subfigure}
    \par
    \begin{subfigure}{0.49\linewidth}
        \centering
        \phantomcaption
        \stackinset{c}{20pt}{t}{15pt}{\subcaptiontext*[3]{}}{\includegraphics[width=\linewidth]{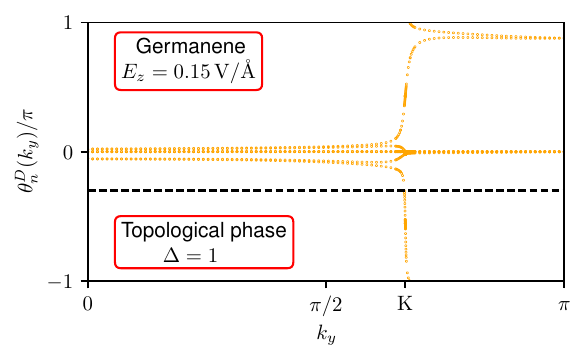}}
        \label{fig:germanene_topo}
    \end{subfigure}
    \hfill
    \begin{subfigure}{0.49\linewidth}
        \centering
        \phantomcaption
        \stackinset{c}{20pt}{t}{15pt}{\subcaptiontext*[4]{}}{\includegraphics[width=\linewidth]{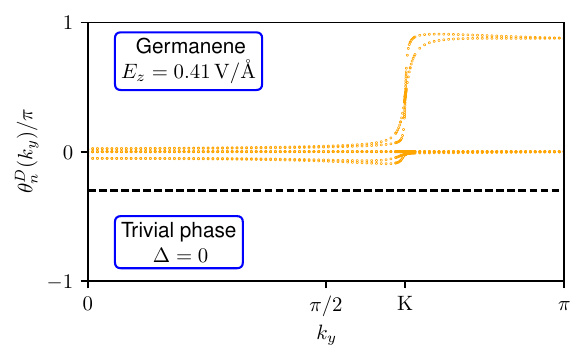}}
        \label{fig:germanene_triv}
    \end{subfigure}
    \par
    \begin{subfigure}{0.49\linewidth}
        \centering
        \phantomcaption
        \stackinset{c}{20pt}{t}{15pt}{\subcaptiontext*[5]{}}{\includegraphics[width=\linewidth]{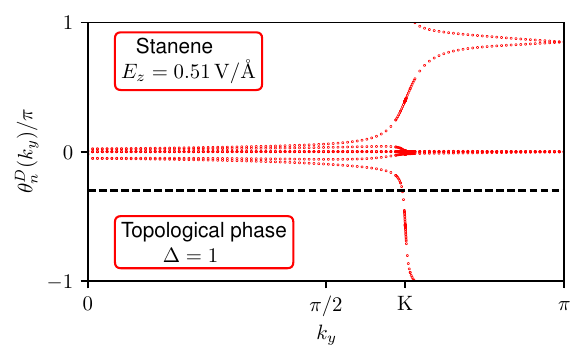}}
        \label{fig:stanene_topo}
    \end{subfigure}
    \hfill
    \begin{subfigure}{0.49\linewidth}
        \centering
        \phantomcaption
        \stackinset{c}{20pt}{t}{15pt}{\subcaptiontext*[6]{}}{\includegraphics[width=\linewidth]{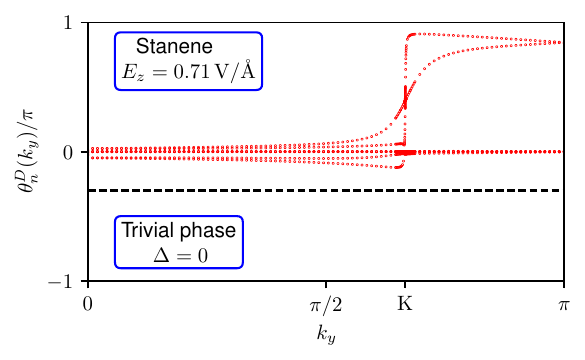}}
        \label{fig:stanene_triv}
    \end{subfigure}
    \caption{Phases $\theta_{n}^{D}(k_y)$ for silicene (top), germanene (middle) and stanene (bottom) in the topological (left) and trivial (right) regimes for the Wannier tight-binding Hamiltonian. An odd number of crossings of an arbitrary reference line $\theta=\theta_0$ across half the Brillouin zone corresponds to a non‑trivial quantum spin Hall phase ($\mathbb{Z}_2=1$) as in \ref{fig:silicene_topo}, \ref{fig:germanene_topo} and \ref{fig:stanene_topo}, whereas an even number corresponds to a trivial insulator ($\mathbb{Z}_2=0$) as in \ref{fig:silicene_triv}, \ref{fig:germanene_triv} and \ref{fig:stanene_triv}.}
    \label{fig:thetas}
\end{figure}

Example plots of the phases $\theta_{n}^{D}$ in terms of $k_y$ are shown in figure \ref{fig:thetas} for silicene, germanene and stanene. In figures \ref{fig:silicene_topo}, \ref{fig:germanene_topo} and \ref{fig:stanene_topo}, the dashed horizontal reference line $\theta=\theta_0$ crosses only one of the evolution lines of the Wannier function centers (odd number), while in figures \ref{fig:silicene_triv}, \ref{fig:germanene_triv} and \ref{fig:stanene_triv} zero evolution lines are crossed (even number). Using equation \eqref{eq:invariant}, this implies that the $\mathbb{Z}_2$ topological invariant is $\Delta = 1$ in the former case and $\Delta = 0$ in the latter, characterizing each of the phases as topological and trivial, respectively.

In addition, the evolution of the Wannier function centers in the topological regime reveals a switching of the Kramer partners between the time-reversal invariant momenta $k_y = 0$ and $k_y = \pi$, as is expected for a topological phase \cite{Fu_2007}. On the other hand, no such switching is observed in the trivial regime. The plots were observed to change from those charactheristic of a topological regime to those indicating a trivial one for a certain critical electric field value $E_z^{\text{cr}}$ in each of the materials considered, implying that the systems experiment a topological phase transition. $E_z^{\text{cr}}$ values, see table \ref{tab:criticalvalues}, were measured at $0.020$ and $\qty{0.250}{V/\angstrom}$ for silicene and germanene, while the result of $\qty{0.690}{V/\angstrom}$ was accurately replicated for stanene \cite{Fuhr_2021}. 

For comparison, we calculate the values at which topological phase transitions are predicted in a four-band tight-binding (TB) model, considering up to second nearest-neighbor interactions \cite{Ezawa_2012_epj}. The Hamiltonian of this model is divided into two parts. The first part consists of three terms inherent to the monolayer ---the nearest-neigbor hopping, the second nearest-neighbor effective spin-orbit coupling (SOC) and the intrinsic Rashba spin-orbit coupling--- which are analysed in depth in \cite{Liu_2011_ham} and are written as follows in the second-quantization formalism
\begin{equation}\label{eq:ham_inherent}
    H_0 =
    -t \sum_{\langle ij \rangle \alpha} c_{i \alpha}^{\dagger} c_{j \alpha}
    +
    i \frac{\lambda_{\mathrm{SO}}}{3\sqrt{3}} \sum_{\langle\!\langle ij \rangle\!\rangle \alpha \beta} \nu_{ij} c_{i \alpha}^{\dagger} \sigma_{\alpha\beta}^{z} c_{j \beta}
    -
    i \frac{2}{3} \lambda_{\mathrm{R}2} \sum_{\langle\!\langle ij \rangle\!\rangle \alpha \beta} \mu_i c_{i \alpha}^{\dagger} (\bm{\sigma} \times \hat{\mathbf{v}}_{ij})^{z}_{\alpha\beta} c_{j \beta}\,,
\end{equation}
where $c_{i \alpha}^{\dagger}$ ($c_{i \alpha}$) is the creation (destruction) operator for an electron in site $i$ with spin $\alpha$ and $\langle ij \rangle$ and $\langle\!\langle ij \rangle\!\rangle$ run thorugh all nearest-neigbor and second-nearest neigbor pairs. $t$ is the transfer energy for the nearest-neighbor hopping, while $\lambda_{\mathrm{SO}}$ and $\lambda_{\mathrm{R}2}$ are the effective and Rashba SOC strenghts, respectively. $\bm{\sigma} = (\sigma_x, \sigma_y, \sigma_z)$ is the Pauli matrix for the spin and $\hat{\mathbf{v}}_{ij} = \mathbf{v}_{ij} / \norm{\mathbf{v}_{ij}}$, with $\mathbf{v}_{ij}$ the vector that connects second-nearest neighbor sites $i$ and $j$. Finally, $\nu_{ij} = +1$ if the atom nearest to the midpoint of the vector that goes from site $i$ to site $j$ lies to the left of said vector as seen from above and $\nu_{ij} = -1$ if it lies to the right, whereas $\mu_{i} = +1$ for the atoms located at $z=\ell$, i.e. a distance $\ell$ upwards with respect to the middle of the monolayer, and $\mu_{i} = -1$ for those located at $z=-\ell$.

When a perpendicular electric field is applied, two other terms are added to the total system Hamiltonian ---an electric potential term and an extrinsic Rashba SOC induced by the applied electric field \cite{Geissler_2013}--- which are
\begin{equation}\label{eq:ham_electric}
    H_E =
    -
    \ell eE_z \sum_{i \alpha} \mu_i c_{i \alpha}^{\dagger} c_{i \alpha}
    +
    i\lambda_{\mathrm{R1}}(E_z) \sum_{\langle ij \rangle \alpha \beta} c_{i \alpha}^{\dagger} (\bm{\sigma} \times \hat{\mathbf{u}}_{ij})^{z}_{\alpha\beta} c_{j \beta}\,,
\end{equation}
where $e$ is the electron charge, $\ell$ is the buckling height, $\lambda_{\mathrm{R}1} (E_z)$ is the strength of the extrinsic Rashba SOC that depends on the intensity of the electric field $E_z$ and $\hat{\mathbf{u}}_{ij} = \mathbf{u}_{ij} / \norm{\mathbf{u}_{ij}}$, with $\mathbf{u}_{ij}$ the vector that connects nearest neighbor sites $i$ and $j$.

The extrinsic Rashba SOC caused by the electric field can be calculated from the expression given in \cite{Geissler_2013}, which is a genaralization of the expression given in \cite{Min_2006} for graphene,
\begin{equation}
    \lambda_{\mathrm{R1}}(E_z) = \frac{e E_z z_0}{3 \sin{(\theta)} V_{sp\sigma}} \xi\,,
\end{equation}
where $e E_z z_0 = - e E_z \abs{\braketOP{n,l=1,m=0}{Z}{n,l=0,m=0}}$ is the first-order nonzero Stark effect matrix element. $z_0$ is given by $-3\sqrt{6}\, a_0/Z^{\ch{Si}}_{\mathrm{eff}}$, $-6\sqrt{5}\, a_0/Z^{\ch{Ge}}_{\mathrm{eff}}$ and $-15\sqrt{2}\, a_0/Z^{\ch{Sn}}_{\mathrm{eff}}$, where $a_0$ is the Bohr radius and $Z_{\mathrm{eff}}$ is the effective nuclear charge for \ch{Si}, \ch{Ge} and \ch{Sn} atoms, respectively, taken from \cite{Clementi_1963} and \cite{Clementi_1967}. In addition, $V_{sp\sigma}$ is the Slater-Koster parameter \cite{Slater_1954} for hopping in the $\sigma$ bond, calculated from \cite{Harrison_1989}, $\theta$ is the buckling angle as defined in \cite{Liu_2011_ham} and $\xi$ is the strength of the atomic spin-orbit interaction. The relevant tight-binding parameters used in this model are shown in table \ref{tab:tbvalues}.

\begin{figure}[thpb]
    \centering
    \includegraphics[width=0.65\textwidth]{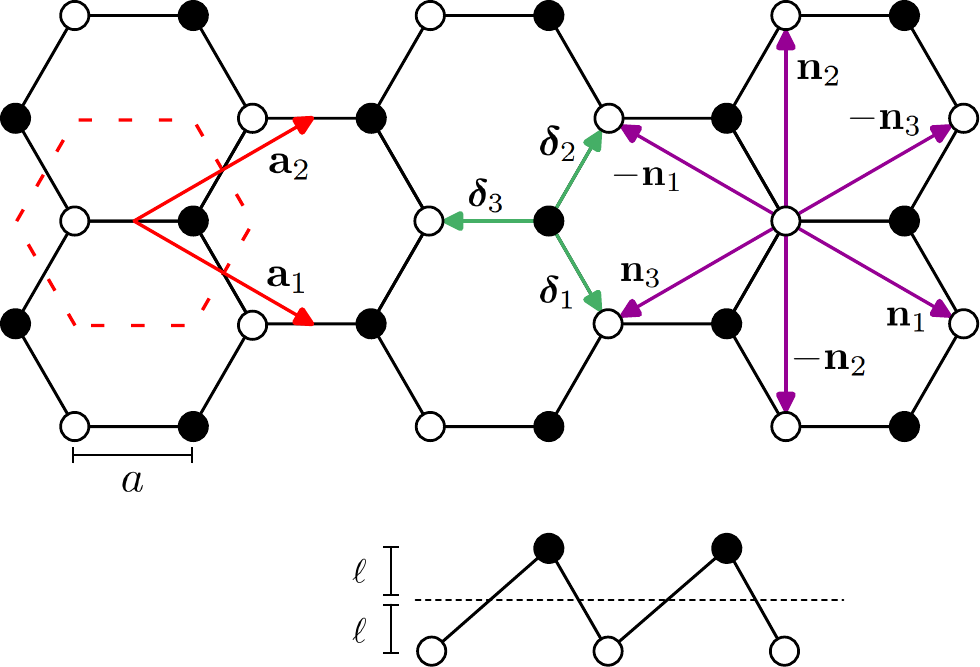}
    \caption{Top (upper) and side (lower) views of a Xene monolayer. Black and white atoms denote sublattices A and B. The red dashed hexagon marks the unit cell; $\mathbf{a}_i$, $\bm{\delta}_i$ and $\mathbf{n}_i$ are lattice, nearest-, and next-nearest-neighbor vectors, respectively. $a$ is the nearest-neighbor distance and $\ell$ is the buckling height.}
    \label{fig:sublattices}
\end{figure}

\ctable[
cap = Comparison of tight-binding values,
caption = {Comparison of relevant tight-binding values for each material based on \cite{Liu_2011_ham}.},
label = tab:tbvalues,
pos = hptb
]{@{}lccccccc@{}}{
\tnote[a]{Slater-Koster parameter calculated from \cite{Harrison_1989}.}
\tnote[b]{Reference \cite{Liu_2011_qshe}.}
\tnote[b]{Reference \cite{Chadi_1977}.}
}{                                                                                 \FL
            & $t (\unit{eV})$ & $\lambda_{\mathrm{SO}} (\unit{meV})$ & $\lambda_{\mathrm{R}2} (\unit{meV})$ &  $V_{sp\sigma}(\unit{eV})$\tmark[a] & $\xi(\unit{eV})$ \ML
Silicene    & $1.65$   & $3.973$   & $0.7$    & $2.81$     & $0.034$\tmark[b] \NN
Germanene   & $1.31$   & $46.3$    & $10.7$   & $2.57$     & $0.196$ \NN
Stanene     & $1.19$   & $64.4$    & $9.5$    & $1.92$   & $0.80$\tmark[c]   \LL
}

The parameters listed in table \ref{tab:tbvalues}, taken from \cite{Liu_2011_ham}, exhibit systematic variations across the Xene series that reflect the underlying atomic properties. The nearest‑neighbour hopping $t$ decreases from silicene to stanene, consistent with the increasing bond length and the more diffuse character of the valence orbitals of the heavier elements. In contrast, the effective spin–orbit coupling $\lambda_{\mathrm{SO}}$ and the atomic SOC strength $\xi$ increase by roughly an order of magnitude, this being the primary factor that allows the existence of a quantum spin Hall phase. The Rashba coupling $\lambda_{\mathrm{R}2}$ exhibits a non‑monotonic variation across the series but remains far smaller than $\lambda_{\mathrm{SO}}$, as expected for the modest buckling heights of the free‑standing layers. The Slater–Koster parameter $V_{sp\sigma}$, computed here from Harrison’s solid‑state table \cite{Harrison_1989}, follows the same gentle decrease with bond length as $t$, consistent with the tight‑binding framework. All parameters are reported explicitly so that the tight‑binding model can be fully reproduced without recourse to supplementary sources.

The total system Hamiltonian is given by $H = H_0 + H_E$, from Equations \eqref{eq:ham_inherent} and \eqref{eq:ham_electric}, and consists of five terms in total. We can express the Hamiltonian in terms of the creation operators ---$a_{i \alpha}^{\dagger}$ and $b_{i \alpha}^{\dagger}$--- and annihilation operators ---$a_{i \alpha}$ and $b_{i \alpha}$--- for the hexagonal $A$ and $B$ sublattices. Using the discrete Fourier transforms of these operators $a_{i \alpha}^{\dagger} = \frac{1}{\sqrt{N}} \sum_{\mathbf{k}}^{\text{B.Z.}} \mathrm{e}^{-\mathrm{i}\mathbf{k}\cdot\mathbf{R}_i} a_{\mathbf{k} \alpha}^{\dagger}$ and $a_{i \alpha} = \frac{1}{\sqrt{N}} \sum_{\mathbf{k}}^{\text{Z.B.}} \mathrm{e}^{\mathrm{i}\mathbf{k}\cdot\mathbf{R}_i} a_{\mathbf{k} \alpha}$, together with the corresponding ones for $b_{i \alpha}^{\dagger}$ and $b_{i \alpha}$, the Hamiltonian in reciprocal space reads
\begin{multline}\label{eq:full_tb_hamiltonian}
    H =
    -t \sum_{\mathbf{k} \alpha}
    \left[
        \phi_{\mathbf{k}} a_{\mathbf{k} \alpha}^{\dagger} b_{\mathbf{k} \alpha}
        +
        \phi_{\mathbf{k}}^* b_{\mathbf{k} \alpha}^{\dagger} a_{\mathbf{k} \alpha}
    \right]
    + \sum_{\mathbf{k} \alpha} \left( i \frac{\lambda_{\mathrm{SO}}}{3\sqrt{3}} \alpha \xi_{\mathbf{k}} - \ell eE_z \right) (a_{\mathbf{k} \alpha}^{\dagger} a_{\mathbf{k} \alpha} - b_{\mathbf{k} \alpha}^{\dagger} b_{\mathbf{k} \alpha})
    \\
    -i\frac{2}{3}\lambda_{\mathrm{R2}} \sum_{\mathbf{k} \alpha \beta}
    \sum_{i=1}^{3} 2i \sin({\mathbf{k}\cdot \mathbf{n}_i})
    \left[a_{\mathbf{k} \alpha}^{\dagger} (\bm{\sigma} \times \hat{\mathbf{n}}_i)_{\alpha \beta}^{z} a_{\mathbf{k} \beta}
    -
    b_{\mathbf{k} \alpha}^{\dagger} (\bm{\sigma} \times \hat{\mathbf{n}}_i)_{\alpha \beta}^{z} b_{\mathbf{k} \beta}\right]
    \\
    +i\lambda_{\mathrm{R1}}(E_z) \sum_{\mathbf{k} \alpha \beta}
    \sum_{i=1}^3
    \left[
        \mathrm{e}^{i\mathbf{k}\cdot \bm{\delta_i}} a_{\mathbf{k} \alpha}^{\dagger} (\bm{\sigma} \times \hat{\bm{\delta}}_i)_{\alpha \beta}^{z} b_{\mathbf{k} \beta}
        -
        \mathrm{e}^{-i\mathbf{k}\cdot \bm{\delta_i}} b_{\mathbf{k} \alpha}^{\dagger} (\bm{\sigma} \times \hat{\bm{\delta}}_i)_{\alpha \beta}^{z} a_{\mathbf{k} \beta}
    \right]\,,
\end{multline}
where the sums in $\mathbf{k}$ run over all vectors in the Brillouin zone, $N$ is the number of unit cells, $\phi_\mathbf{k}=\sum_{i=1}^3 \mathrm{e}^{i\mathbf{k}\cdot\bm{\delta}_i}$ and $\xi_{\mathbf{k}} = -2i \sum_{i=1}^3 \sin(\mathbf{k}\cdot\mathbf{n}_i)$, with $\bm{\delta}_i$ and $\mathbf{n}_i$ the vectors to first and second neighbors, respectively, as defined in figure \ref{fig:sublattices}.

The energy spectrum calculated from the above Hamiltonian consists of four bands, which at point $K=\frac{4\pi}{3a}(\frac{\sqrt{3}}{2},\frac{1}{2})$ is given by
\begin{equation}
    \begin{gathered}
        E_{1,2} = \pm\lambda_{\mathrm{SO}} - \ell eE_z\\
        E_{3,4} = \lambda_{\mathrm{SO}} \pm\sqrt{(\ell eE_z)^2 + (3\lambda_{\mathrm{R1}}(E_z))^2}
    \end{gathered}
\end{equation}

\begin{figure}[thpb]
    \centering
     \begin{subfigure}{0.49\linewidth}
        \centering
        \phantomcaption
        \stackinset{l}{3mm}{t}{3mm}{\captiontext*{}}{\includegraphics[width=\linewidth]{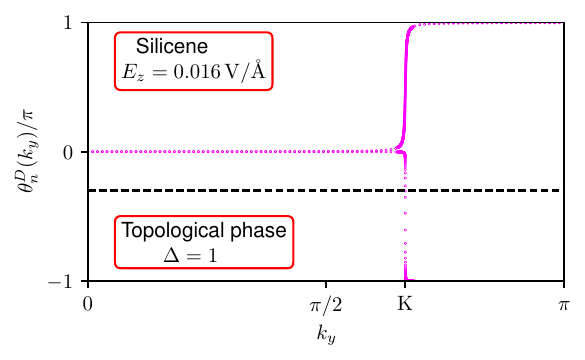}}
        \label{fig:silicene_topo_tb}
    \end{subfigure}
    \hfill
    \begin{subfigure}{0.49\linewidth}
        \centering
        \phantomcaption
        \stackinset{l}{3mm}{t}{3mm}{\captiontext*{}}{\includegraphics[width=\linewidth]{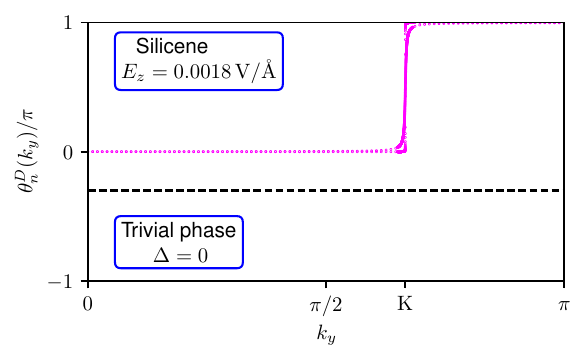}}
        \label{fig:silicene_triv_tb}
    \end{subfigure}
    \par
    \begin{subfigure}{0.49\linewidth}
        \centering
        \phantomcaption
        \stackinset{c}{20pt}{t}{15pt}{\subcaptiontext*[3]{}}{\includegraphics[width=\linewidth]{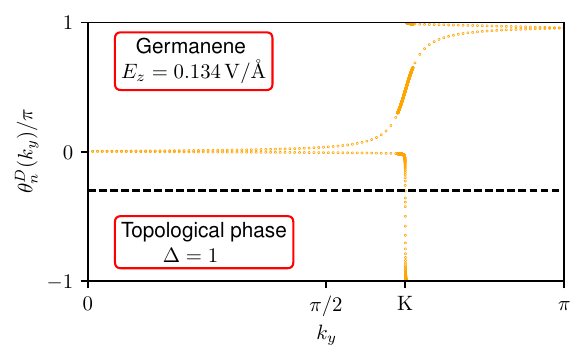}}
        \label{fig:germanene_topo_tb}
    \end{subfigure}
    \hfill
    \begin{subfigure}{0.49\linewidth}
        \centering
        \phantomcaption
        \stackinset{c}{20pt}{t}{15pt}{\subcaptiontext*[4]{}}{\includegraphics[width=\linewidth]{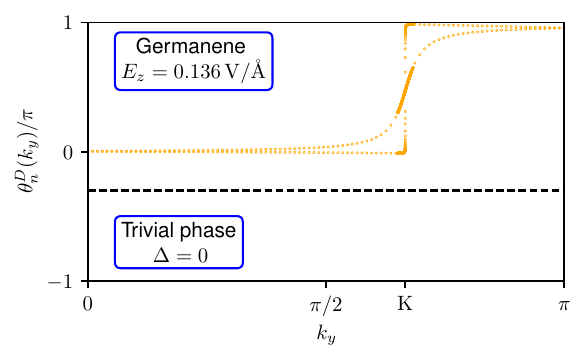}}
        \label{fig:germanene_triv_tb}
    \end{subfigure}
    \par
    \begin{subfigure}{0.49\linewidth}
        \centering
        \phantomcaption
        \stackinset{c}{20pt}{t}{15pt}{\subcaptiontext*[5]{}}{\includegraphics[width=\linewidth]{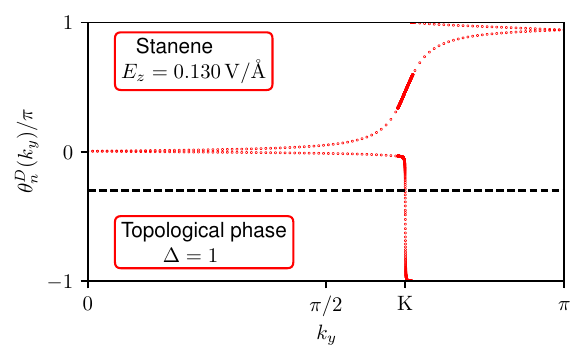}}
        \label{fig:stanene_topo_tb}
    \end{subfigure}
    \hfill
    \begin{subfigure}{0.49\linewidth}
        \centering
        \phantomcaption
        \stackinset{c}{20pt}{t}{15pt}{\subcaptiontext*[6]{}}{\includegraphics[width=\linewidth]{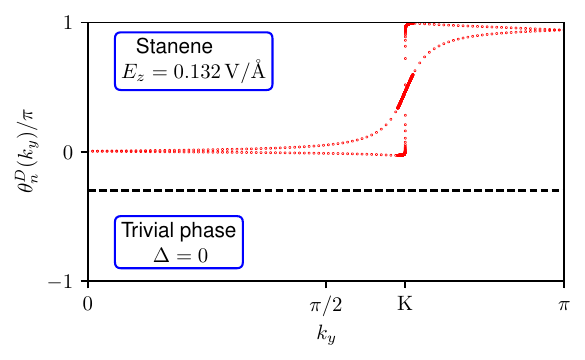}}
        \label{fig:stanene_triv_tb}
    \end{subfigure}
    \caption{Phases $\theta_{n}^{D}(k_y)$ for silicene (top), germanene (middle) and stanene (bottom) in the topological (left) and trivial (right) regimes for the four-band next-nearest-neighbor tight-binding model in \eqref{eq:full_tb_hamiltonian}. In this model, the topological phase transitions occur at $\qty{0.0174}{eV/\angstrom}$, $\qty{0.135}{eV/\angstrom}$ and $\qty{0.132}{eV/\angstrom}$, much lower critical electric field values than those predicted from the Wannierisation-based model.}
    \label{fig:thetas_tb}
\end{figure}

The closing of the direct bandgap at point $\mathrm{K}$ gives the condition for calculating the critical electric field for the topological phase transitions
\begin{equation}
    E_z^{\text{cr}}
    =
    \frac{\lambda_{\mathrm{SO}}}{\ell}
    \left[
        1 - \left(
                    \frac{3\lambda_{\mathrm{R1}(E_z)}}{2 \lambda_{\mathrm{SO}}}
            \right)^2
    \right]\,,
\end{equation}
where it can be seen that only the extrinsic Rashba effect term modifies the $\frac{\lambda_{\mathrm{SO}}}{\ell}$ value obtained using the Kane and Mele Hamiltonian without Rashba terms \cite{Kane_2005}. With the values presented in table \ref{tab:tbvalues} this correction is of the order of $\sim \num{e-4}$, $\sim \num{e-3}$ and $\sim \num{e-2}$ for silicene, germanene and stanene, respectively, giving critical electric field values which are shown in table \ref{tab:criticalvalues}. For stanene, the extrinsic Rashba term substantially lowers the $E_z^{\text{cr}}$ value compared to the values expected without the Rashba effect: $\qty{0.0175}{eV/\angstrom}$ for silicene, $\qty{0.136}{eV/\angstrom}$ for germanene, and $\qty{0.150}{eV/\angstrom}$ for stanene. For all three materials studied, the values obtained using the tight-binding model underestimate the values obtained by first-principles calculations, with the largest difference obtained in the case of stanene. To confirm that these bandgap closings at these $E_z^{\text{cr}}$ values were indeed indicative of topological phase transitions, the $\mathbb{Z}_2$ invariant was calculated using the method outlined previously in section \ref{section:Z2}. Sample plots for electric field values below and above the phase transitions are depicted in figure \ref{fig:thetas_tb} for all the three materials studied.

\ctable[
cap = Comparison of critical electric field values,
caption = {Comparison of critical electric field values in $\unit{V/\angstrom}$.},
label = tab:criticalvalues,
pos = hptb
]{@{}lcccc@{}}{
\tnote[a]{Values obtained from first-principles calculations.}
\tnote[b]{Values reported in \cite{Yu_2018}.}
\tnote[c]{Values obtained from the tight-binding model.}
}{                                                                         \FL
                                       & Silicene  & Germanene  & Stanene  \ML
\tmark[a]Calculated $E_z^{\text{cr}}$  & $0.020$   & $0.250$    & $0.690$  \NN
\tmark[b]Reference $E_z^{\text{cr}}$   & $0.004$   & $0.04$     & $0.23$   \NN
\tmark[c]TB model $E_z^{\text{cr}}$    & $0.0174$   & $0.135$    & $0.132$  \LL
}

As we can see from the comparison in table \ref{tab:criticalvalues}, the method used in the present work of sequentially using DFT and Wannierisation followed by a posterior calculation of the topological $\mathbb{Z}_2$ invariant for each of the values $E_z$ of electric field considered, allows obtaining critical field values $E_z^{\text{cr}}$ that are more precise than those obtained based on a tight-binding model fitted only with first-principles data of the system with no electric field applied ---second row--- or solely from tight-binding approximations ---third row---, both of which result in an underestimation of $E_z^{\text{cr}}$.

As noted in the Introduction, free-standing group‑14 Xenes are, experimentally, invariably stabilised by a supporting substrate. The interaction with the substrate typically breaks the sublattice symmetry and introduces charge transfer and orbital hybridisation, which can strongly modify the electronic structure. Consequently, measured gaps for supported Xenes differ substantially from their expected intrinsic values.

Experimental efforts to measure the bandgap of silicene have focused extensively on growth on \ch{Ag}(111) substrates, yielding diverse results. Using Angle-Resolved Photoemission Spectroscopy (ARPES), Vogt suggested that silicene on \ch{Ag}(111) possessed a Dirac-like band structure modified by substrate interactions with a gap opening of approximately \qty{0.6}{eV} \cite{Vogt_2012}. In contrast, Scanning Tunneling Spectroscopy (STS) studies reported a much smaller superconducting gap of \qty{35}{meV} \cite{Chen_2013}. Meanwhile, other ARPES investigations in \cite{DePadova_2013} into multilayer silicene islands reported gapless Dirac fermions. These findings have faced significant criticism from works such as \cite{Tsoutsou_2013, Mahatha_2014, Mahatha_2015, Shirai_2014, Mannix_2014}, which contend that these electronic signatures do not represent intrinsic silicene but instead originate from substrate-modified silver interface states or represent diamond-like $sp^3$ silicon films terminated by silver surface reconstructions.

On the other hand, silicene nanosheets grown on highly oriented pyrolytic graphite (HOPG) were found in \cite{DeCrescenzi_2016} via STS to exhibit metallic character, effectively having a zero bandgap. On \ch{ZrB2}(0001), \cite{Fleurence_2012} reported the opening of a direct $\pi$-electronic band gap through ARPES, attributing this gap to substrate-induced buckling. These conflicting measured values underscore the persistent challenge of isolating the intrinsic electronic properties of silicene from the heavy influence of the supporting substrate. The free‑standing results presented here therefore provide an essential baseline for disentangling the intrinsic topological properties from substrate‑driven effects, a separation that is crucial for future engineering of supported topological phases.

Experimental measurements of the bandgap for germanene have consistently yielded values indicating metallic or gapless behavior, falling short of theoretically predicted spin-orbit gaps. Initial STS studies of germanene grown on \ch{Ge2Pt} clusters (on a \ch{Ge}(110) substrate) reported metallic-like behavior \cite{Bampoulis_2014}. Subsequent STS investigations on the same system at room temperature identified a V-shaped density of states, a signature of a 2D Dirac material, but without a detectable gap \cite{Zhang_2015,}. Similarly, STS measurements of germanene on \ch{MoS2} \cite{Zhang_2016} and the second layer of germanene islands on \ch{Cu}(111) \cite{Qin_2017} showed no evidence of a spin-orbit gap. Additionally, STS measurements at cryogenic temperatures (\qty{77}{K}) for germanene on \ch{Ge2Pt} clusters concluded that any intrinsic spin-orbit gap must be smaller than \qty{6}{meV}, as the observed rounding of the Dirac point could be explained by thermal broadening \cite{Walhout_2016}. These collective results are critically reviewed in \cite{Castenmiller_2020}, which argues that the missing gap in STS experiments is due to the electric field generated in the tunnel junction. This field, caused by the work function difference between germanene ($\sim \qty{3.8}{eV}$) and typical STM tips ($\qtyrange{4.5}{5.5}{eV}$), is believed to suppress the intrinsic spin-orbit gap.

Experimental investigations have identified stanene as gapless or metallic, largely due to strong substrate interactions that obscure its intrinsic electronic properties. The first realization of 2D stanene via molecular beam epitaxy on a \ch{Bi2Te3}(111) substrate, reported in \cite{Zhu_2015}, found the material to be metallic. ARPES measurements in this study showed that the $p_z$ bands theoretically expected to define the gap near the K-point had completely disappeared from the spectra and the authors argued that the $p_z$ orbitals were likely fully saturated by interactions with the underlying substrate or the presence of adsorbates like hydrogen, effectively rendering the identification of a topological gap impossible. In contrast, on a \ch{Cu}(111) substrate, ARPES measurements in \cite{Deng_2018} identified a spin-orbit coupling (SOC)-induced bandgap of \qty{0.3}{eV}. This value was observed in what the authors described as ``ultra-flat stanene'', where interaction with the copper substrate supposedly suppressed the intrinsic buckling of the tin layer. Another significant non-zero measurement was reported for stanene grown on \ch{Sb}-terminated \ch{InSb}(111), where ARPES revealed a bandgap of \qty{0.44}{eV} \cite{Xu_2018}. This larger gap was attributed to the specific interfacial environment and strain provided by the \ch{InSb} template, making it a promising candidate for room-temperature QSH applications.

Epitaxial synthesis is almost exclusively done on metallic substrates which inherently screen any applied perpendicular electric field, making in situ gating and simultaneous ARPES/STS measurements very difficult. Nevertheless, Bampoulis et al. reported the experimental realization of epitaxial germanene grown on \ch{Ge2Pt}(101) as a buckled-honeycomb quantum spin hall (QSH) insulator, altering the topological state of germanene using the built-in electric field in the tip-sample tunneling junction \cite{Bampoulis_2023}. The authors used a first layer of germanium that acts as a buffer, electronically decoupling the subsequent germanene monolayer from the bulk metallic substrate beneath it and allowing germanene to maintain its intrinsic quantum spin Hall properties without the substrate overpowering them. Rather than trying to transfer the highly reactive germanene to a dielectric substrate to build a traditional Field Effect Transistor (FET), they used the STM tip itself as a local, nanoscale gate by controlling the tip-sample separation distance and metal coating on the tip. They found germanene starts with a remarkably large bulk band gap of roughly \qty{70}{meV} at lower fields (\qty{0.16}{V/\angstrom}). This is significantly larger than theoretical predictions for freestanding germanene ($\sim \qty{24}{meV}$, see table \ref{tab:dftvalues}), which the authors attribute to a complex interplay of stacking, strain, and proximity effects with the substrate. As they increased the electric field strength, the topological gap steadily decreased until it completely closed at a critical electric field of \qty{0.195}{V/\angstrom}, transitioning into a gapless Dirac semimetal. Pushing the electric field beyond the critical point resulted in a trivial band gap opening, accompanied by the complete disappearance of the conductive edge states. While early theoretical models for freestanding layers predict an electric-field-induced gap closure at roughly \qty{0.04}{V/\angstrom} \cite{Yu_2018}, these values deviate significantly from experimental reality due to the neglect of field-induced changes in the electronic structure. By incorporating these effects, our calculations yield a critical field of \qty{0.250}{V/\angstrom} for germanene. This represents a substantial improvement over prior theoretical predictions, substantially closer to the experimentally obtained value of \qty{0.195}{V/\angstrom} in \cite{Bampoulis_2023} than prior estimates \qty{0.04}{V/\angstrom} \cite{Yu_2018}, with a residual $\sim \qty{28}{\percent}$ deviation attributable to the substrate-induced enhancement discussed above.

\section{Conclusions}\label{section:conclusions}
In this work, we have systematically investigated the topological phase transitions in the group-14 Xenes under the influence of a perpendicular electric field. By combining density-functional theory with Wannier-based Hamiltonians and a gauge-independent computation of the $\mathbb{Z}_2$ topological invariant, we used a fully \textit{ab initio} and self-consistent workflow capable of accurately describing the evolution of the electronic structure and topology across the transition. This approach allowed us to clearly identify two distinct insulating regimes in all studied materials: a non-trivial topological insulator phase characterized by time-reversal symmetry-protected edge states, and a trivial insulating phase emerging beyond the critical field strength. The methodology offers a significant advance over previous tight-binding approaches, which often underestimate the critical electric field due to their limited treatment of field-induced band structure modifications. By explicitly accounting for these changes at each field strength prior to constructing the Wannier tight-binding Hamiltonian, our results provide more reliable and quantitatively precise estimates of the transition thresholds $E_z^{\text{cr}}$, namely, $\num{0.020}$, $\qty{0.250}{V/\angstrom}$ and $\qty{0.690}{V/\angstrom}$ for silicene, germanene and stanene, thus improving the predictive capacity of theoretical models in this domain. These features are essential for the rational design of next-generation devices that exploit electric-field-controlled topological phase transitions including topological transistors, spintronic components, and quantum-information architectures. Future work will extend this approach to plumbene, other 2D families (e.g., group‑15/‑16 Xenes, MXenes, and graphdiyne), more complex van der Waals heterostructures and chemically engineered 2D systems, aiming to discover materials with tunable topological responses and enhanced potential for technological applications.

\section*{Funding}
J. A. Villarreal Murúa acknowledges funding from Agencia Nacional de Investigación e Innovación (ANII) through a postdoctoral grant (code \nolinkurl{PD\_NAC\_2024\_1\_182525}), which made this research possible.


\printbibliography

\end{document}